\documentclass[10pt, final, journal, letterpaper, twocolumn]{IEEEtran}

\usepackage[dvips]{graphicx}
\usepackage{times}
\usepackage{cite}
\usepackage{amsmath}
\usepackage{array}
\usepackage{amssymb}

\usepackage{bbm}

\usepackage{stfloats}
\usepackage{rotating,threeparttable,booktabs}
\usepackage{bm}
\usepackage{dcolumn,booktabs}
\usepackage{multirow}
\usepackage{graphicx}
\usepackage{subfigure}
\usepackage{color}

\usepackage[letterpaper]{geometry}

\usepackage{amsmath,amsthm,amssymb,amsfonts}
\makeatletter
\thm@headfont{\sc}
\makeatother

\begin{document}
\IEEEoverridecommandlockouts
\title{\huge{Beamforming for Simultaneous Wireless Information and Power Transfer in Two-Way Relay Channels}\vspace{-10pt}}
\author{\IEEEauthorblockN{Wei~Wang$^{*}$, Rui~Wang$^{\dagger}$, and Hani Mehrpouyan$^{\ddagger}$}\\
\IEEEauthorblockA{$^{*}$Department of Communication Engineering,~Nantong University, Jiangsu, China.\\
$^{\dagger}$Department of Information and Communications,~Tongji University, Shanghai, China.\\
$^{\ddagger}$Department of Electrical and Computer Engineering, Boise State University, Boise, USA.\\
Emails: wwang2011@ntu.edu.cn, ruiwang@tongji.edu.cn, hanimehrpouyan@boisestate.edu.}\vspace{-27pt}
\thanks{This work is supported by the National Natural Science Foundation of China under grant 61401313, 61371113, the Natural Science Foundation of Jiangsu Province under grant BK20130393, the Project of Flagship-Major Construction of Jiangsu Higher Education Institutions of China under grant 06150032/002 and NASA Early Investigation Grant.}}
\maketitle
\vspace{-1cm}
\begin{abstract}

This paper studies simultaneous wireless information and power transfer (SWIPT) systems in two-way relaying (TWR) channels. Here, two source nodes receive information and energy simultaneously via power splitting (PS) from the signals sent by a multi-antenna relay node. Our objective is to maximize the weighted sum of the harvested energy at two source nodes subject to quality of service (QoS) constraints and the relay power constraints. Three well-known and practical two-way relay strategies are considered, i.e., amplify-and-forward (AF), bit level XOR based decode-and-forward (DF-XOR) and symbol level superposition coding based DF (DF-SUP). For each relaying strategy, we formulate the joint energy transmit beamforming and PS ratios optimization as a nonconvex quadratically constrained problem. To find a closed-form solution of the formulated problem, we decouple the primal problem into two subproblems. In the first problem, we intend to optimize beamforming vector for a given PS ratio. In the second subproblem, we optimize the PS ratio with a given beamforming vector. It is worth noting that although the corresponding subproblem are nonconvex, the optimal solution of each subproblem can still be found by using certain techniques. We provide numerical results that demonstrate the advantage of adapting the different relaying strategies and weighted factors to harvest energy in two-way relaying channel.

\end{abstract}


\section{Introduction}

Energy harvesting (EH) from surrounding environments is an emerging solution to prolong the operational time of energy-constrained nodes in wireless networks \cite{Ulukus2015,Bi2015}. Compared with conventional energy sources, radio frequency (RF) signals can carry both information and energy simultaneously. Simultaneous wireless information and power transfer (SWIPT) has recently drawn significant attention, where SWIPT has been investigated for various wireless channels\cite{Varshney2008,Grover2010,Liu2013,Zhang2013,Park2013,Xu2014,Shi2014,Huang2013,Zhou2014}. For example, a point-to-point single-antenna additive white Gaussian noise (AWGN) channel was first studied in \cite{Varshney2008}, where the authors used a capacity-energy function to study the fundamental performance tradeoff for simultaneous information and power transfer. Later on, SWIPT was extended to a frequency selective channels in \cite{Grover2010}. The authors in \cite{Liu2013} studied SWIPT for fading channels subject to time-varying co-channel interference. In \cite{Zhang2013,Park2013}, SWIPT schemes for multiple-input-multiple-output (MIMO) channels were considered. The transmit beamforming design was studied for SWIPT in multiple-input-single-output (MISO) broadcast channels in \cite{Xu2014,Shi2014}. Moreover, SWIPT has been investigated in other physical layer setups such as the OFDM, and more. in \cite{Huang2013,Zhou2014}.

Besides the above studies related to one-hop transmission, SWIPT technique has also been extended to wireless relay networks\cite{Krikidis2013,Nasir2013,Gurakan2013,Zeng2015,Chen2014,Li2013,Li2014,Wen2014,Tutuncuoglu2015}. For the one-way single-antenna relay channel, two protocols, namely time switching (TS) and power splitting (PS), are proposed for amplify-and-forward (AF) relay networks in \cite{Krikidis2013,Nasir2013}. Later on, SWIPT was extended to a full-duplex wireless-powered one-way relay channel in \cite{Gurakan2013,Zeng2015}, where the data and energy queues of the relay are updated simultaneously in every time slot. Because two-way relaying (TWR) is able to simultaneously enlarge wireless coverage and enhance spectral efficiency, the SWIPT protocols for TWR channel recently have attracted much attention. In \cite{Chen2014}, the authors provided a SWIPT protocol in two-way AF relaying channels, where two sources exchange information via an energy harvesting relay node. The authors investigated the sum-rate maximization problem in two-way AF relaying channels in \cite{Li2013}, where two source nodes harvest energy from multiple relay nodes. In \cite{Li2014}, The authors studied the relay beamforming design problem for SWIPT in a non-regenerative two-way multi-antenna relay network. The authors investigated a compute-and-forward (CF) relay networks optimal beamforming design problem in \cite{Wen2014}, where two source nodes harvest energy by SWIPT from relay nodes. Moreover, for different relaying strategies, the authors in \cite{Tutuncuoglu2015} studied the sum-throughput maximization problem in a two-way AWGN relay channel, where all nodes are powered by EH.

\subsection{Motivation}

\begin{figure}[t]
\begin{centering}
\includegraphics[scale=0.63]{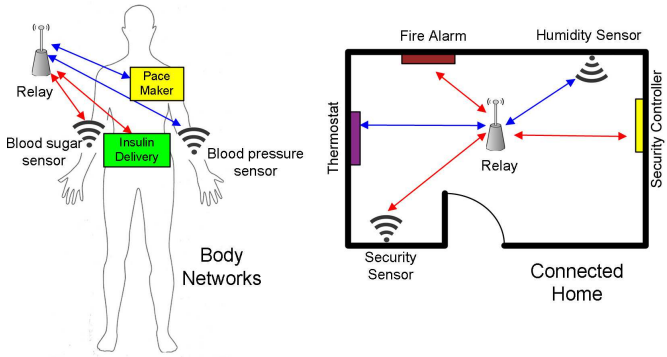}
\vspace{-0.7cm}
\caption{Example applications of SWIPT in TWR systems with battery-limited sensor nodes.} \label{application}
\end{centering}
\vspace{-0.5cm}
\end{figure}

So far, most studies on SWIPT in relay networks focused on energy-constrained relay nodes\cite{Krikidis2013,Nasir2013,Gurakan2013,Zeng2015,Chen2014}, \cite{Tutuncuoglu2015}. As a matter of fact, the sensor nodes or other low-power devices often have very limited battery storage and require an external charging sources to remain active in wireless cooperative or sensor networks. Although replacing or recharging batteries provides a solution to this problem, it may incur a high cost and sometimes even be unavailable due to some physical or economic limitations. As show in Fig.~\ref{application}, when the sensor node that is inside the body or embedded in a building structure is depleted of energy, it cannot fulfill its role any longer unless the source of energy is replenished. Therefore, EH in such kind of scenarios, where the relay node serves as the energy source, is particularly important as it can provide a much safer and much more convenient solution. Hence, in this paper, we consider a TWR SWIPT system with battery-limited source nodes and a relay node that acts also as a source of energy (as described with more detail in Section II). Under this setup, the source nodes receive information and energy simultaneously from the signals sent by a relay node. Furthermore, to enhance bandwidth efficiency and power transfer, we consider a scenario, where the relay node is equipped with multiple antennas. This setup applies to lots of practical wireless transmission scenarios. Since TS can be regarded as a special case of PS with only binary split power ratios \cite{Liu2013,Zhang2013}, we focus our study on PS receivers instead of TS receivers.

\subsection{Related Works}
To the best of our knowledge, only three papers, e.g., \cite{Li2013}, \cite{Li2014} and \cite{Wen2014}, thus far studied multi-antenna TWR SWIPT systems with the battery-limited source nodes. In \cite{Li2013} and \cite{Li2014}, the authors considered TWR SWIPT systems and studied the optimal relay beamforming problem which maximizes the weighted sum rate subject to the transmit power constraint at relay and the EH constraint at source nodes. However, the authors in \cite{Li2013} assumed that the source node is able to decode information and extract power simultaneously, which, as explained in \cite{Zhang2013}, may not hold in practice. In contrast to \cite{Li2013}, the authors in \cite{Li2014} considered the case of separated EH and information decoding (ID) receivers. The energy receiver could harvest energy from the signals transmitted by both the source and the relay, and the source node can only receive the information from the signals forwarded by the relay. Furthermore, in their works, only AF relaying strategy was considered. Authors in \cite{Wen2014} considered a similar overlay protocol as ours. However, it focused on the relay transmit power minimization problem for a TWR SWIPT network with CF relaying strategy. In this paper, we study a TWR based PS-SWIPT system where TWR consists of a multi-antenna relay node and two single-antenna source nodes. Here, two source nodes receive information and energy simultaneously via PS from the signals sent by the relay node. In particular, we consider three types relaying strategies: AF, bit level XOR based decode-and-forward (DF-XOR) and symbol level superposition coding based DF (DF-SUP). Different from \cite{Li2013,Li2014,Wen2014}, our objective is to maximize the weighted sum of the harvested energy at two source nodes subject to a given minimum signal-to-interference-and-noise ratio (SINR) constraint at source nodes and a maximum transmit power constraint at the relay node. This scenario is of particular interest in supporting sensor networks that empower the next set of applications such as internet of things and also medical monitoring as outlined in Fig.~\ref{application}. In these applications, a large number of sensors will be operating in close vicinity. Hence, SINR is a important metric for maintaining a given throughput while maximizing energy transfer of the sensors or source nodes by the relay. The latter maximizes the operational time of the sensors which can be a important metric in the scenarios presented in Fig.~\ref{application}. To the authors' best knowledge, the joint beamforming and PS optimization for this new setup has not been studied in existing works.

\subsection{Our Contributions}
Under the above setup, a TWR based PS-SWIPT system is considered in this paper. Different from existing works, we assume that source nodes can receive information and energy simultaneously via PS from the signals sent by the relay node. Moreover, various two-way relaying strategies may result in different transmit signals at the relay node. The impact of various relaying strategies on the amount of harvest energy has not been considered in existing studies. Besides, for another challenging doubly-near-far problem \cite{Bi2015} in the TWR SWIPT system, which refers to a node far away from the relay harvests much lower energy but consumes more to transmit data than a node near the relay, could be mitigated effectively by setting different EH priorities for different source nodes.

The main contributions of this work are summarized as follows. \emph{Firstly}, this is the first work to investigate joint beamforming and PS optimization for a TWR SWIPT system with battery-limited source nodes. Here, two source nodes receive information and energy simultaneously via PS from the signals sent by a multi-antenna relay node. To achieve this goal, we propose a two-phase PS-based relaying protocol. \emph{Secondly}, based on the above system setup, we present different transmit signals at the relay node by considering three practical two-way relaying strategies due to their implementation simplicity\cite{Laneman2003, Laneman2004}, i.e., AF, DF-XOR and DF-SUP. To explore the performance limit of the system , for each relaying strategy, we formulate the joint energy transmit beamforming and PS ratios optimization as a nonconvex quadratically constrained problem. \emph{Thirdly}, for each nonconvex relay beamforming optimization problem, we find a solution by decoupling the primal problem into two subproblems. The first subproblem only optimizes the beamforming vectors. We solve this nonconvex problem by applying the technique of semidefinite programming (SDP)\cite{Huang2010}. The second subproblem only includes the PS ratios. We propose an novel algorithm to find the optimal closed-form solutions by separating the latter nonconvex subproblem into eight cases. Then, a near optimal solution of the original optimization problem is found based on a two-tie iterative algorithm. \emph{Finally}, we provide numerical results for each relaying scheme to evaluate the performance of the proposed optimal beamforming designs. It is shown that when the priority and the distance of two source nodes are symmetric, the DF-XOR relaying strategy performs better than the other two strategies. While the distances of two source nodes are asymmetric, for all three considered relaying strategies, the furthest node can harvest more energy when its energy weight factor is set to a larger value, which can provide an effective solution to the doubly-near-far problem\cite{Bi2015}.

\subsection{Organization}
The remainder of this paper is organized as follows. The TWR SWIPT system model is described in Section II. In Section III, the weighted sum-power harvested maximization problems are formulated for different relaying strategies. The solutions for the associated optimization problems by using suitable optimization tools are presented in Section IV. In Section V, numerical simulation results are provided. Finally, the paper is concluded in Section VI.

\emph{Notations}:
Boldface lowercase and uppercase letters denote vectors and matrices, respectively.
For a square matrix ${\bf A}$, ${\bf A}^T$, ${\bf A}^{*}$, ${\bf A}^H$, ${\rm Tr}({\bf A})$, ${\rm Rank}(\bf A)$ and $||{\bf A}||$ denote its transpose, conjugate, conjugate transpose, trace, rank, and Frobenius norm, respectively. ${\bf A}\succeq 0$ indicates that ${\bf A}$ is a positive semidefinite matrix.
${\rm vec}({\bf A})$ denotes the vectorization operation by stacking the columns of ${\bf A}$ into a single vector ${\bf a}$.
$\mathbb E(\cdot)$ denotes the statistical expectation.
$\otimes$ denotes the Kronecker product.
$\oplus$ denotes the XOR operator.
${\bf 0}$ and ${\bf I}$ denote the zero and identity matrix, respectively.
The distribution of a circular symmetric complex Gaussian vector with mean vector $\bf x$ and covariance matrix ${\bf \Sigma} $ is denoted by ${\cal CN}({\bf x},{\bf \Sigma})$.
${\mathbb C}^{x \times y}$ denotes the $x \times y$ domain of complex matrices.


\section{System Model} \vspace{-1pt}

\begin{figure}[t]
\begin{centering}
\includegraphics[scale=0.62]{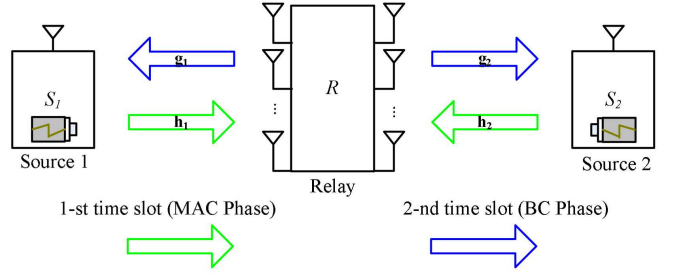}
\vspace{-0.7cm}
\caption{A two-time slot TWR system, where each source node coordinates information decoding and energy harvesting.} \label{SystemModel}
\end{centering}
\vspace{-0.5cm}
\end{figure}

Consider a half-duplex TWR system where two single-antenna source nodes $S_1$ and $S_2$ exchange information with each other through an $N$-antenna relay node, $R$, as shown in Fig.~\ref{SystemModel}. The channel matrices from $S_1$ and $S_2$ to the relay are denoted by ${\bf h}_1$ and ${\bf h}_2$, respectively, and the channel matrices from the relay to $S_1$ and $S_2$ are denoted by ${\bf g}_1$ and ${\bf g}_2$, respectively. To further improve the spectral efficiency, the two-time slot TWR model is used to realize bidirectional communication. Throughout this paper, the following set of assumptions are made:

$\bullet$ The source nodes cannot communication with each other directly. Hence, all messages are sent through the relay. This occurs when the direct link is blocked due to long-distance path loss or obstacles \cite{Hasna2002,Xiong2014}.

$\bullet$ The relay is connected to the power grid, which implies that it has access to reliable power at all times. However, the source nodes are powered by the energy limited batteries or capacitors, and need to replenish their energy by wireless power transfer.

$\bullet$ Amongst the different relaying protocols, AF, DF-XOR and DF-SUP schemes are applied at the relay node due to their implementation simplicity \cite{Laneman2003, Laneman2004}.


$\bullet$ Quasi-static block fading channels are assumed here, i.e., channels are unchanged in a time-slot of $T$, but change from time slot to time slot. The use of such channels is motivated by prior research in this field \cite{Liu2013,Zhang2013,Park2013,Xu2014,Shi2014}, \cite{Zeng2015,Chen2014,Li2013} and practical consideration.


\begin{figure}[t]
\begin{centering}
\includegraphics[scale=0.68]{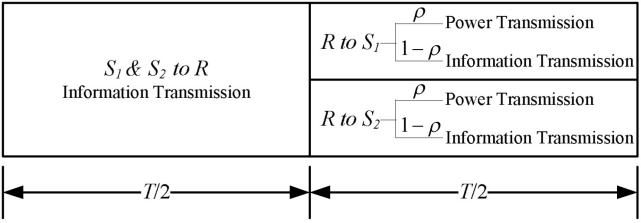}
\vspace{-0.3cm}
\caption{Energy harvesting and information processing relaying protocol based on PS with splitting ratio $\rho$.} \label{protocol}
\end{centering}
\vspace{-0.5cm}
\end{figure}

As shown in Fig.~\ref{protocol}, we propose a two-phase PS-based protocol for the TWR system. In the first phase of duration $T/2$, two source nodes $S_1$ and $S_2$ deliver their information to the relay node $R$ simultaneously. In the second phase with the remaining time duration $T/2$, the received information signal at $R$ is processed by the aforementioned relaying strategies and then forwarded to the source nodes. Note that here, by assuming a PS ratio, $\rho$, the transmit signal from the relay is used to simultaneously achieve information and power transfer.

Based on the above system setup, the received signal at the relay after the first phase, i.e., \emph{the multiple access (MAC) phase}, is given by \vspace{-3pt}
\begin{equation}\label{EQU-1} \vspace{-3pt}
{\bf y}_R = {\bf h}_1 x_1 + {\bf h}_2 x_2  + {\bf n}_R,
\end{equation}
where $x_i$, for $i\in \{1,2 \}$, represents the transmit signal from node $S_i$, ${\bf h}_i \in \mathbb{C}^{N \times 1}$ is the channel vector from node $S_i$ to the relay node $R$, and ${\bf n}_R$ denotes the additive complex Gaussian noise vector at the relay following ${\cal CN}({\bf 0},{\bf \sigma}^2_r I_N)$. Each transmit signal $x_i$ is assumed to satisfy an average power constraint, i.e., $\mathbb E(|x_i|^2)=P_i$.

Upon receiving ${\bf y}_R$, the relay node performs certain processing and then forwards its signal to the source nodes in the second phase, also
referred as \emph{broadcast (BC) phase}. Let the transmit signal from the relay be denoted by
\begin{equation}\label{EQU-2} \vspace{-3pt}
{\bf x}_R = {\bf x}_{12}  + {\bf x},
\end{equation}
where ${\bf x}_{12}$ is the combined signal consisting of the messages from two nodes by using physical-layer network coding (PLNC). Note that, here, besides ${\bf x}_{12}$,
we also include a new signal ${\bf x}$, which provides us with more degrees of freedom to optimize power transfer from relay to the source nodes.

It is worth noting that various two-way relaying strategies may result in different transmit signal ${\bf x}_R$. The three relaying strategies we considered, namely, AF, DF-XOR and DF-SUP are all favorable for practical implementation and the precoding designs based on these strategies are mathematically tractable \cite{Laneman2003, Laneman2004}. The primary focus of this work is to maximize the weighted sum of the harvested power based on practical two-way relay strategies, while meeting a minimum SINR for each source node.

\section{Relaying Strategies and Optimization Problem Formulations} \vspace{-1pt}
Based on the channel setup described in Section II, in this section we shall present different transmit signals ${\bf x}_R$ for the TWR SWIPT system by considering three practical two-way
relay strategies. Moreover, to explore the system performance limit, we also formulate three optimization problems for these approaches in this section.

\subsection{AF Relay Strategy}
With the AF relaying strategy, the relay transmit signal ${\bf x}_R$ in \eqref{EQU-2} can be expressed as
\begin{equation}\label{EQU-3} \vspace{-3pt}
{\bf x}_R = {\bf x}_{12} + {\bf x}={\bf W} {\bf h}_1 x_1 + {\bf W} {\bf h}_2 x_2  + {\bf W} {\bf n}_R+ {\bf x},
\end{equation}
where ${\bf W}$ represents the precoding matrix used at the relay. In addition, we assume that the relay node has the maximum transmit power $P_r$, i.e., ${\rm Tr}\{\mathbb E({\bf x}_R {\bf x}_R^H)\}\leq P_r$, which is equivalent to
\begin{equation}\label{EQU-4} \vspace{-3pt}
P_1 ||{\bf W}{\bf h_1}||^2_2 + P_2 ||{\bf W}{\bf h_2}||^2_2+{\rm Tr}({\bf Q}_x)+{\sigma}^2_r ||{\bf W}||^2_F\leq P_r,
\end{equation}
where ${\bf Q}_x= \mathbb E({\bf x} {\bf x}^H)$ is the covariance matrix of ${\bf x}$, and $P_1$ and $P_2$ are the transmit powers used at nodes $S_1$ and $S_2$ , respectively. Then, the received signals at the two nodes in the second $T/2$ time interval are given by
\begin{equation}\label{EQU-5} \vspace{-3pt}
\begin{split}
{\tilde{y}_i}={\bf g}^T_i {\bf W} {\bf h}_{\overline{i}} {\tilde{x}_{\overline{i}}}+&{\bf g}^T_i {\bf W} {\bf h}_i {\tilde{x}_i}+{\bf g}^T_i {\tilde{\bf x}}+{\bf g}^T_i {\bf W} {\tilde{\bf n}_R}+{n}_{i,d},
\end{split}
\end{equation}
where ${\overline{i}}=2$ if $i=1$ and ${\overline{i}}=1$ if $i=2$. Note that here ${\tilde{y}_i}$, ${\tilde{x}_i}$, ${\tilde{\bf x}}$ and ${\tilde{\bf n}_R}$ denote the signals in the RF band and $n_{i,d}$ is the additive Gaussian noise due to the receiving antenna that follows ${\cal CN}({\rm 0},{\bf \sigma}^2_{i,d})$ for $i\in \{1,2 \}$. Specifically, as shown in Fig.~\ref{protocol}, the received signal ${\tilde{y}_i}$ at each end node is split into two portions for EH and information processing. Let $\rho\in (0,1 )$ be the power splitting ratio, meaning that $\sqrt {1-\rho} {\tilde{y}_i}$ is used for information processing. As a result, after converting the received signal to baseband and performing self-interference cancelation, the obtained signal is denoted as
\begin{equation}\label{EQU-6} \vspace{-3pt}
{y_i}={\sqrt {1-\rho}}({\bf g}^T_i {\bf W} {\bf h}_{\overline{i}} x_{\overline{i}}+{\bf g}^T_i{\bf x}+{\bf g}^T_i {\bf W} {\tilde{\bf n}_R}+{n}_{i,d})+{n}_{i,c},
\end{equation}
where $n_{i,c} \sim {\cal CN}({\rm 0},{\bf \sigma}^2_{i,c})$ is the additive Gaussian noise introduced by the signal conversion from RF band to baseband. Accordingly, the SINR at the node $S_i$ is given by
\begin{equation}\label{EQU-7} \vspace{-3pt}
{\rm SINR}^{AF}_i={\frac{P_{\overline{i}} |{\bf g}^T_i {\bf W} {\bf h}_{\overline{i}}|^2}{{\bf g}^T_i{\bf Q}_x {\bf g}^*_i+{\bf \sigma}^2_r ||{\bf g}^T_i {\bf W}||^2_2+{\bf \sigma}^2_{i,d}+{\frac{{\bf \sigma}^2_{i,c}}{1-\rho}}}}.
\end{equation}
Moreover, the other portion of the received signal, $\sqrt {\rho} {\tilde{y}_i}$, is used for EH. Since the background noise at the EH receiver is negligible and thus can be ignored \cite{Zhang2013}, the harvested energy, $E_i$ during EH time $T/2$ is given by
\begin{equation}\label{EQU-8} \vspace{-3pt}
{E_i}={\frac{\eta T}{2}}{\rho}(|{\bf g}^T_i {\bf W} {\bf h}_{\overline{i}}|^2 P_{\overline{i}}+|{\bf g}^T_i {\bf W} {\bf h}_i|^2 P_i+{\bf g}^T_i{\bf Q}_x {\bf g}^*_i),
\end{equation}
where $\eta$ is the energy conversion efficiency with $0<\eta<1$ which depends on the rectification process and the EH circuitry \cite{Zhang2013}. Note that in \eqref{EQU-8}, the self-interference can be used for EH, which is different from information processing.

Our design goal is to maximize the weighted sum power harvested at two EH nodes, which is defined as the harvested energy minus the consumed energy. The corresponding optimization problem can be formulated as
\begin{equation}\label{EQU-9} \vspace{-3pt}
\begin{split}
&\max_{ P_1,P_2,{\rho},{\bf W},{\bf Q}_x \succeq 0} ~~ \alpha(E_1-\frac{P_1 T}{2})+\beta(E_2-\frac{P_2 T}{2})\\
\textit{s.t.} ~~ &{\rm SINR}^{AF}_i \geq \tau_i,~i=1,2,\\
&P_i \leq P_{max,i},~i=1,2,\\
&{\rm Tr}\{\mathbb E({\bf x}_R {\bf x}_R^H)\}\leq P_r.
\end{split}
\end{equation}
In \eqref{EQU-9}, $\alpha$ and $\beta$ correspond to the given energy weights for the two EH receivers $S_1$ and $S_2$, respectively, where a larger weight value indicates a higher priority of transferring energy to the corresponding EH receiver as compared to other EH receiver. $\tau_i$ and $P_{max,i}$ are the SINR requirement and the maximum transmit power at node $S_i$, respectively.

\subsection{DF-XOR Relay Strategy}
If the relay node adopts the DF relaying strategy, it needs to decode the messages sent from both source nodes in the MAC phase, and then transmit a function of the two messages in the BC phase. The rate region of the MAC channel is characterized by \cite{Oechtering2008,Wang2013}
\begin{equation}\label{EQU-10} \vspace{-3pt}
\begin{split}
C_{MAC}({\tilde{R}_1},{\tilde{R}_2})=\quad\quad\quad\quad\quad\quad\quad\quad\quad\quad\quad\quad\quad\\
\begin{cases}
\tilde{R}_1 \leq \log_2(1+\frac{P_1||{\bf h}_1||^2_2}{{\bf \sigma}^2_r})\\
\tilde{R}_2 \leq \log_2(1+\frac{P_2||{\bf h}_2||^2_2}{{\bf \sigma}^2_r})\\
\tilde{R}_1+\tilde{R}_2 \leq \log_2 \det({\bf I}+{\frac{P_1}{{\bf \sigma}^2_r}}{\bf h}_1{\bf h}^H_1+{\frac{P_2}{{\bf \sigma}^2_r}}{\bf h}_2{\bf h}^H_2),
\end{cases}
\end{split}
\end{equation}
where $\tilde{R}_1$ and $\tilde{R}_2$ are the transmit rates at nodes $S_1$ and $S_2$, respectively.

We assume that the messages sent from the nodes in the MAC phase can be successfully decoded at the relay node. Let ${\bf b}_i$ denote the decoded bit sequence from ${S}_i$,
for $i\in \{1,2 \}$. With the DF-XOR relaying strategy, the combined bit sequence is yielded as ${\bf b}_{12}={\bf b}_1 \oplus {\bf b}_2$.
Then the transmit signal in the second time interval of $T/2$, denoted by ${\bf x}_R$ in \eqref{EQU-2}, can be expressed as
\begin{equation}\label{EQU-11} \vspace{-3pt}
{\bf x}_R = {\bf s}_{12}  + {\bf x},
\end{equation}
where ${\bf s}_{12}$ is the modulated signal of bit sequence ${\bf b}_{12}$. The corresponding relay power constraint for signal ${\bf x}_R$ is denoted as
\begin{equation}\label{EQU-12} \vspace{-3pt}
{\rm Tr}\{\mathbb E({\bf x}_R {\bf x}_R^H)\}={\rm Tr}({\bf Q}_s)+{\rm Tr}({\bf Q}_x)\leq P_r,
\end{equation}
where ${\bf Q}_s= \mathbb E({\bf s}_{12} {\bf s}^H_{12})$ is the covariance matrix of ${\bf s}_{12}$. Then, the received signals at node $S_i$ in the RF band is given by
\begin{equation}\label{EQU-13} \vspace{-3pt}
\begin{split}
{\tilde{y}_i}={\bf g}^T_i \tilde{\bf s}_{12}+{\bf g}^T_i \tilde{\bf x}+{n}_{i,d},  {i=1,2.}
\end{split}
\end{equation}
Using power splitting, $\sqrt {1-\rho} {\tilde{y}_i}$ is used for information processing at the end nodes. After baseband conversion and self-interference cancelation, the obtained signal is denoted as
\begin{equation}\label{EQU-14} \vspace{-3pt}
{y_i}={\sqrt {1-\rho}}({\bf g}^T_i {\bf s}_{12}+{\bf g}^T_i {\bf x}+{n}_{i,d})+{n}_{i,c},
\end{equation}
Subsequently, the SINR at the node $S_i$ can be determined as
\begin{equation}\label{EQU-15} \vspace{-3pt}
{\rm SINR}^{XOR}_i=\frac{{\bf g}^T_i {\bf Q}_s {\bf g}^*_i}{{\bf g}^T_i {\bf Q}_x {\bf g}^*_i+{\bf \sigma}^2_{i,d}+\frac{{\bf \sigma}^2_{i,c}}{1-\rho}}.
\end{equation}

On the other hand, $\sqrt {\rho} {\tilde{y}_i}$ is used for EH at the node $S_i$. The harvested energy, $E_i$ is given as \cite{Zhang2013}
\begin{equation}\label{EQU-16} \vspace{-3pt}
{E_i}={\frac{\eta T}{2}}{\rho}({\bf g}^T_i {\bf Q}_s {\bf g}^*_i+{\bf g}^T_i {\bf Q}_x {\bf g}^*_i), {i=1,2.}
\end{equation}

Similarly, our aim is to maximize the weighted sum power harvested at the two source nodes subject to a given minimum SINR constraint at each source node and a maximum transmit power constraint at the relay node. The corresponding optimization problem can be formulated as
\begin{equation}\label{EQU-17} \vspace{-3pt}
\begin{split}
&\max_{{\rho},{\bf Q}_s\succeq 0,{\bf Q}_x \succeq 0} ~~ \alpha(E_1-\frac{P_1 T}{2})+\beta(E_2-\frac{P_2 T}{2})  \\
\textit{s.t.} ~~ &{\rm SINR}^{XOR}_i \geq \tau_i,~i=1,2,\\
&{\rm Tr}({\bf Q}_s)+{\rm Tr}({\bf Q}_x)\leq P_r.
\end{split}
\end{equation}
Note that here different from \eqref{EQU-9}, it is not necessary to optimize ${P}_1$ and ${P}_2$ as they are determined via the constraints presented in \eqref{EQU-10}.

\subsection{DF-SUP Relay Strategy}
In this subsection, we consider a case where the relay uses the DF-SUP relaying strategy. Again, we assume that the messages sent from the nodes in the MAC phase can be successfully decoded at the relay node. Then by applying DF-SUP relaying strategy, the transmit signal in the BC phase, denoted by ${\bf x}_R$, can be expressed as
\begin{equation}\label{EQU-18} \vspace{-3pt}
{\bf x}_R = {\bf s}_1 +{\bf s}_2+ {\bf x},
\end{equation}
where ${\bf s}_i$ is the modulated signal of bit sequence of node $S_i$. The corresponding relay power constraint for signal
${\bf x}_R$ is denoted as
\begin{equation}\label{EQU-19} \vspace{-3pt}
{\rm Tr}\{\mathbb E({\bf x}_R {\bf x}_R^H)\}={\rm Tr}({\bf Q}_{s,1})+{\rm Tr}({\bf Q}_{s,2})+{\rm Tr}({\bf Q}_x)\leq P_r,
\end{equation}
where ${\bf Q}_{s,i}= \mathbb E({\bf s}_i {\bf s}^H_i)$ is the covariance matrix of ${\bf s}_i$. Then, the received signals at node $S_i$ in the RF band is given by
\begin{equation}\label{EQU-20} \vspace{-3pt}
\begin{split}
{\tilde{y}_i}={\bf g}^T_i \tilde{\bf s}_1+{\bf g}^T_i \tilde{\bf s}_2+{\bf g}^T_i \tilde{\bf x}+{n}_{i,d},  {i=1,2.}
\end{split}
\end{equation}
Let us assume that $\sqrt {1-\rho} {\tilde{y}_i}$ portion of the received signal is used for information processing at the end nodes. After converting this signal to baseband and performing self-interference cancelation, the obtained signal is given by
\begin{equation}\label{EQU-21} \vspace{-3pt}
{y_i}={\sqrt {1-\rho}}({\bf g}^T_i {\bf s}_{\overline{i}}+{\bf g}^T_i {\bf x}+{n}_{i,d})+{n}_{i,c},
\end{equation}
Then, the SINR at the node $S_i$ can be denoted as
\begin{equation}\label{EQU-22} \vspace{-3pt}
{\rm SINR}^{SUP}_i=\frac{{\bf g}^T_i {\bf Q}_{s,{\overline{i}}} {\bf g}^*_i}{{\bf g}^T_i {\bf Q}_x {\bf g}^*_i+{\bf \sigma}^2_{i,d}+\frac{{\bf \sigma}^2_{i,c}}{1-\rho}}.
\end{equation}

On the other hand, $\sqrt {\rho} {\tilde{y}_i}$ is used for EH at the node $S_i$. The harvested energy, $E_i$, is given as \cite{Zhang2013}
\begin{equation}\label{EQU-23} \vspace{-3pt}
{E_i}={\frac{\eta T}{2}}{\rho}({\bf g}^T_i {\bf Q}_{s,1} {\bf g}^*_i+{\bf g}^T_i {\bf Q}_{s,2} {\bf g}^*_i+{\bf g}^T_i {\bf Q}_x {\bf g}^*_i),
\end{equation}

Also, our design goal is to maximize the weighted sum power harvested at the two EH nodes, which is defined as the harvested energy minus the consumed energy. The corresponding optimization problem can be formulated as
\begin{equation}\label{EQU-24} \vspace{-3pt}
\begin{split}
&\max_{{\rho},{\bf Q}_{s,1}\succeq 0,{\bf Q}_{s,2}\succeq 0,{\bf Q}_x \succeq 0} ~~ \alpha(E_1-\frac{P_1 T}{2})\\
&\quad\quad\quad\quad\quad\quad\quad\quad\quad\quad+\beta(E_2-\frac{P_2 T}{2})  \\
\textit{s.t.} ~~ &{\rm SINR}^{SUP}_i \geq \tau_i,~i=1,2,\\
&{\rm Tr}({\bf Q}_{s,1})+{\rm Tr}({\bf Q}_{s,2})+{\rm Tr}({\bf Q}_x)\leq P_r.
\end{split}
\end{equation}
Here, similar to \eqref{EQU-17}, we are not necessary to optimize ${P}_1$ and ${P}_2$ as they are determined by \eqref{EQU-10}.

For different relaying strategies, we have formulated three weighted sum-power harvested maximization problems in \eqref{EQU-9}, \eqref{EQU-17} and \eqref{EQU-24}. In the following sections, we will propose three algorithms to solve these optimization problems.

\section{Optimal Design of Three Maximization Problems} \vspace{-1pt}
For the AF relaying strategy, the optimization problem in \eqref{EQU-9} is nonconvex due to not only the coupled beamforming vectors $\{\bf W, {\bf Q}_x \}$ and the remaining parameters $\{P_i, \rho \}$ in both the SINR and transmitted power constraints but also all the quadratic terms involving ${\bf W}$. In general, it is difficult or even intractable to obtain the global optimal solution to a nonconvex problem \cite{Shi2014,Li2014}. However, it is well known that a function can be maximized by first maximizing over some of the variables, and then maximizing over the remaining ones [28, Sec 4.1.3]. Thus, when $P_i$ and $\rho$ are first fixed, the resulting beamforming optimization problem reduces to that of a conventional nonconvex problem with a rank-one constraint. The latter can be efficiently solved by using some rank relaxation techniques \cite{Huang2010}. Moreover, we note that when the beamforming vectors $\{\bf W, {\bf Q}_x \}$ are fixed, the resulting beamforming optimization problem over $\{P_i, \rho \}$ is still a nonconvex problem. However, as show later the optimal solution can be obtained in closed-form by separating this subproblem into eight cases. In the following, we first decouple problem \eqref{EQU-9} into two subproblems that can be solved separately, and then propose a two-tie iterative algorithm to obtain the near optimal solution of the original optimization problem. Finally, similarly, we decouple problems \eqref{EQU-17} and \eqref{EQU-24} into two subproblems. Note that, here, different from \eqref{EQU-9}, two subproblems from \eqref{EQU-17} and \eqref{EQU-24} only involve the beamforming vectors and PS ratios.

\subsection{Joint Beamforming and PS Optimization for AF relaying strategy} \vspace{-1pt}
Let us solve the two subproblems stemming from \eqref{EQU-9}. In the first subproblem, we try to find the solutions of ${\bf W}$ and ${\bf Q}_x$ for fixed ${P}_1$, ${P}_2$ and ${\rho}$ values. Then, we update the values of ${P}_1$, ${P}_2$ and ${\rho}$ by fixing the remaining parameters. These subproblems are updated in an alternating manner.

\textit{1) Optimize ${\bf W}$ and ${\bf Q}_x$ for fixed ${P}_1$, ${P}_2$ and ${\rho}$}: Note that when fixing ${P}_1$, ${P}_2$ and ${\rho}$, the problem of optimizing variables ${\bf W}$ and ${\bf Q}_x$ is equivalent to
\begin{equation}\label{EQU-25} \vspace{-3pt}
\begin{split}
&\max_{{\bf W},{\bf Q}_x \succeq 0}~~\alpha\rho(|{\bf g}^T_1 {\bf W} {\bf h}_2|^2 P_2+|{\bf g}^T_1 {\bf W} {\bf h}_1|^2 P_1\\
&\quad\quad\quad\quad\quad+{\bf g}^T_1 {\bf Q}_x {\bf g}^*_1 )+ \beta\rho(|{\bf g}^T_2 {\bf W} {\bf h}_1|^2 P_1\\
&\quad\quad\quad\quad\quad+|{\bf g}^T_2 {\bf W} {\bf h}_2|^2 P_2+{\bf g}^T_2 {\bf Q}_x {\bf g}^*_2 ) \\
\textit{s.t.} ~~ &{\rm SINR}^{AF}_i\geq {\tau}_i,~i=1,2.\\
&{\rm P_1} ||{\bf W}{\bf h}_1||^2_2 + {\rm P_2} ||{\bf W}{\bf h}_2||^2_2+{\rm Tr}({\bf Q}_x)\\
&+{\bf \sigma}^2_r ||{\bf W}||^2_F\leq P_r,
\end{split}
\end{equation}
Although problem \eqref{EQU-25} has a simpler form than the original problem in \eqref{EQU-9}, it is still a nonconvex problem. To find the optimal solution of problem \eqref{EQU-25}, we conduct some further transformations on \eqref{EQU-25}. To be specific, we transform $|{\bf g}^T_1 {\bf W} {\bf h}_2|^2$ and $||{\bf W}{\bf h}_1||^2_2$ into their equivalent forms as
\begin{align}
{|{\bf g}^T_1 {\bf W} {\bf h}_2|^2}&= {\rm Tr}({\bf g}^T_1 {\bf W} {\bf h}_2 {\bf h}^H_2 {\bf W}^H {\bf g}^*_1)\tag{26a}\\
&={\rm Tr}({\bf g}^*_1 {\bf g}^T_1 {\bf W} {\bf h}_2 {\bf h}^H_2 {\bf W}^H)\tag{26b}\\
&={\bf w}^H ({\bf h}^*_2 {\bf h}^T_2 \otimes {\bf g}^*_1 {\bf g}^T_1){\bf w}\tag{26c}\label{EQU-26c}\\
&={\rm Tr}(({\bf h}^*_2 {\bf h}^T_2 \otimes {\bf g}^*_1 {\bf g}^T_1){\bf w}{\bf w}^H),\tag{26d}
\end{align}
and similarly
\begin{align}
{||{\bf W}{\bf h}_1||^2_2}&= {\rm Tr}({\bf W}{\bf h}_1{\bf h}^H_1{\bf W}^H)\tag{27a}\\
&={\rm Tr}({\bf I}{\bf W}{\bf h}_1{\bf h}^H_1{\bf W}^H)\tag{27b}\\
&={\bf w}^H ({\bf h}^*_1 {\bf h}^T_1 \otimes {\bf I}){\bf w}\tag{27c}\label{EQU-27c}\\
&={\rm Tr}(({\bf h}^*_1 {\bf h}^T_1 \otimes {\bf I}){\bf w}{\bf w}^H).\tag{27d}
\end{align}
where ${\bf w}={\rm vec}({\bf W})$. In obtaining \eqref{EQU-26c} and \eqref{EQU-27c}, we have used the identity
\setcounter{equation}{27}
\begin{align}
{\rm Tr}({\bf ABCD})={({\rm vec}({\bf{D}}^T))^T}({{\bf C}^T} \otimes \bf{A}){\rm vec}({\bf{B}}).
\end{align}

Similar to (26) and (27), we apply the above transformations to other terms in \eqref{EQU-25}. Let ${\tilde{\bf W}}\triangleq{\bf w}{\bf w}^H$, \eqref{EQU-25} can be rewritten as
\begin{equation}\label{EQU-29} \vspace{-3pt}
\begin{split}
&\max_{{\tilde{\bf W}}\succeq 0,{\bf Q}_x \succeq 0}~~{\rm Tr}({\bf A}_1 {\tilde{\bf W}})+{\rm Tr}({\bf B}_1 {\bf Q}_x)\\
\textit{s.t.} ~~&{\rm Tr}({\bf C}^i_1 {\tilde{\bf W}})-{\rm Tr}({\tau}_i{\bf g}^*_i {\bf g}^T_i {\bf Q}_x)\geq D^i_1,~i=1,2.\\
&{\rm Tr}({\bf E}_1{\tilde{\bf W}})+{\rm Tr}({\bf Q}_x)\leq P_r,\\
&{\rm Rank({\tilde{\bf W}})}=1.
\end{split}
\end{equation}
where ${\bf A}_1\triangleq{(P_2 {\bf h}^*_2 {\bf h}^T_2+P_1 {\bf h}^*_1 {\bf h}^T_1)} \otimes (\alpha\rho{\bf g}^*_1 {\bf g}^T_1+\beta\rho{\bf g}^*_2 {\bf g}^T_2)$,
${\bf B}_1\triangleq{(\alpha\rho{\bf g}^*_1 {\bf g}^T_1+\beta\rho{\bf g}^*_2 {\bf g}^T_2)}$,
${\bf C}^i_1 \triangleq {(P_{\overline{i}} {\bf h}^*_{\overline{i}} {\bf h}^T_{\overline{i}}-{\tau}_i{\bf \sigma}^2_r {I})} \otimes {{\bf g}^*_i {\bf g}^T_i}$,
${D}^i_1\triangleq({\bf \sigma}^2_{i,d}+{\frac{{\bf \sigma}^2_{i,c}}{1-\rho}}){\tau}_i$, and ${\bf E}_1\triangleq{(P_1 {\bf h}^*_1 {\bf h}^T_1+P_2 {\bf h}^*_2 {\bf h}^T_2+{\bf \sigma}^2_r {I})}\otimes {I}$.
Due to the rank-one constraint, finding the optimal solution of \eqref{EQU-29} is difficult. We therefore resort to relaxing it by deleting the rank-one constraint, namely,
\begin{equation}\label{EQU-30} \vspace{-3pt}
\begin{split}
&\max_{{\tilde{\bf W}}\succeq 0,{\bf Q}_x \succeq 0}~~{\rm Tr}({\bf A}_1 {\tilde{\bf W}})+{\rm Tr}({\bf B}_1 {\bf Q}_x)\\
\textit{s.t.} ~~&{\rm Tr}({\bf C}^i_1 {\tilde{\bf W}})-{\rm Tr}({\tau}_i{\bf g}^*_i {\bf g}^T_i {\bf Q}_x)\geq D^i_1,~i=1,2.\\
&{\rm Tr}({\bf E}_1{\tilde{\bf W}})+{\rm Tr}({\bf Q}_x)\leq P_r,\\
\end{split}
\end{equation}
It is noted that if the problem \eqref{EQU-30} has a rank-one optimal solution of ${\tilde{\bf W}}$, i.e., the problem \eqref{EQU-29} has a rank-one optimal solution of ${\tilde{\bf W}}$, the problem \eqref{EQU-29} is equivalent to the problem \eqref{EQU-25}. Fortunately, we have the following lemma.

\textit{Lemma 1}: The rank-one optimal solution of the problem \eqref{EQU-30} always exists.
{\proof} Please refer to Appendix A.

By acquiring the optimal rank-one solution of \eqref{EQU-30}, we can further get the optimal solution of \eqref{EQU-29} and then the optimal solution of \eqref{EQU-25}.

\textit{2) Optimize ${P}_1$, ${P}_2$ and ${\rho}$ for fixed ${\bf W}$ and ${\bf Q}_x$}: In the second step, we need to optimize the power ${P}_1$, ${P}_2$ and the power ratio
${\rho}$ with the remaining variables fixed. The corresponding optimization problem can be formulated as
\begin{equation}\label{EQU-31} \vspace{-3pt}
\begin{split}
&\max_{P_1,P_2,{\rho}} ~~ \alpha(E_1-\frac{P_1 T}{2})+\beta(E_2-\frac{P_2 T}{2})\\
\textit{s.t.} ~~ &{\rm SINR}^{AF}_i\geq {\tau}_i,~i=1,2.\\
&{P_1} ||{\bf W}{\bf h}_1||^2_2 + {P_2} ||{\bf W}{\bf h}_2||^2_2+{\rm Tr}({\bf Q}_x)\\
&+{\bf \sigma}^2_r ||{\bf W}||^2_F\leq P_r,\\
&{0< P_i \leq P_{max,i}},~i=1,2.\\
&{0<\rho<1},
\end{split}
\end{equation}
Similar to (26) and (27), we apply the transformations in \eqref{EQU-31}. The problem of optimizing the variables ${P}_1$, ${P}_2$ and ${\rho}$ is equivalent to
\begin{align}
&\max_{{P}_1,{P}_2,{\rho}} ~~ {A_2 \rho P_2+B_2 \rho P_1 -\alpha P_1-\beta P_2+C_2 \rho}\tag{32a}\label{32a}\\
\textit{s.t.} ~~&{(E_2P_2-D_2)(1-\rho)}\geq {\tau}_1{\bf \sigma}^2_{1,c},\tag{32b}\label{32b} \\
&{(G_2P_1-F_2)(1-\rho)}\geq {\tau}_2{\bf \sigma}^2_{2,c}, \tag{32c}\label{32c}\\
&{P_1 J_2+P_2 K_2}\leq P_r-L_2,\tag{32d}\label{32d}\\
&0<P_1\leq {P}_{max,1},\tag{32e}\label{32e}\\
&0<P_2\leq {P}_{max,2},\tag{32f}\label{32f}\\
&{0<\rho<1}.\tag{32g}\label{32g}
\end{align}
where $A_2\triangleq{\frac{\alpha\eta T}{2}} |{\bf g}^T_1 {\bf W} {\bf h}_2|^2+{\frac{\beta\eta T}{2}} |{\bf g}^T_2 {\bf W} {\bf h}_2|^2$, $B_2\triangleq{\frac{\alpha\eta T}{2}} |{\bf g}^T_1 {\bf W} {\bf h}_1|^2+{\frac{\beta\eta T}{2}} |{\bf g}^T_2 {\bf W} {\bf h}_1|^2$, $C_2\triangleq{\frac{\alpha\eta T}{2}} {\bf g}^T_1 {\bf Q}_x {\bf g}^*_1+{\frac{\beta\eta T}{2}} {\bf g}^T_2 {\bf Q}_x {\bf g}^*_2$, $D_2\triangleq({\bf g}^T_1 {\bf Q}_x {\bf g}^*_1+{\bf \sigma}^2_r ||{\bf g}^T_1{\bf W}||^2_2+{\bf \sigma}^2_{1,d}){\tau}_1$, $E_2\triangleq|{\bf g}^T_1 {\bf W} {\bf h}_2|^2$, $F_2\triangleq({\bf g}^T_2 {\bf Q}_x {\bf g}^*_2+{\bf \sigma}^2_r ||{\bf g}^T_2{\bf W}||^2_2+{\bf \sigma}^2_{2,d}){\tau}_2$, $G_2\triangleq|{\bf g}^T_2 {\bf W} {\bf h}_2|^2$, $J_2\triangleq||{\bf W} {\bf h}_1||^2_2$, $K_2\triangleq||{\bf W} {\bf h}_2||^2_2$ and $L_2\triangleq{\rm Tr}({\bf Q}_x)+{\bf \sigma}^2_r ||{\bf W}||^2_F$.

Problem (32) is still nonconvex in its current form since both the SINR and transmit power constraints involve coupled transmit power ${P}_i$ and power ratio ${\rho}$. To find the optimal solution of (32), we give the following lemma.

\textit{Lemma 2}\label{lemma2}: Let $\{P^*_1, P^*_2, \rho^* \}$ denote an optimal solution of problem (32), we have\\
(1) for the optimal solution $\{P^*_1, P^*_2, \rho^* \}$, either the SINR or the transmit power constraint must hold with equality;\\
(2) the optimal solution $\{P^*_1, P^*_2, \rho^* \}$ can be obtained in closed-form by comparing the following eight cases:
\begin{itemize}
\item When the two SINR constraints \eqref{32b} and \eqref{32c} hold with equality, the optimal solution $\{P^*_1, P^*_2, \rho^* \}$ are given by\\
\setcounter{equation}{32}
\begin{equation}\label{EQU-33} \vspace{-3pt}
\begin{split}
&P^*_1 = {\frac{\frac{{\tau}_2{\bf \sigma}^2_{2,c}}{1-\rho^*}+F_2}{G_2}}, \quad P^*_2 = {\frac{\frac{{\tau}_1{\bf \sigma}^2_{1,c}}{1-\rho^*}+D_2}{E_2}},\\ &{\rho^*}=1-\sqrt{\frac{a_1+a_2-a_3}{a_1}}.
\end{split}
\end{equation}
where $a_1\triangleq-(A_2D_2G_2+B_2E_2F_2+C_2E_2G_2)$, $a_2\triangleq A_2E_2{\tau}_1{\bf \sigma}^2_{1,c}+B_2E_2{\tau}_2{\bf \sigma}^2_{2,c}+A_2D_2G_2$ and $a_3\triangleq\alpha E_2{\tau}_2{\bf \sigma}^2_{2,c}+\beta G_2{\tau}_1{\bf \sigma}^2_{1,c}$.

\item When the constraints \eqref{32b} and \eqref{32d} hold with equality, the optimal solution $\{P^*_1, P^*_2, \rho^* \}$ are given by\\
\begin{equation}\label{EQU-34} \vspace{-3pt}
\begin{split}
\hspace{0.5cm}&P^*_1  = {\frac{P_r-L_2-P^*_2K_2}{J_2}}, \hspace{0.15cm}P^*_2 = {\frac{\frac{{\tau}_1{\bf \sigma}^2_{1,c}}{1-\rho^*}+D_2}{E_2}},\\ &{\rho^*}=1-\sqrt{-\frac{b_1+b_2}{J_2E_2b_3}}.
\end{split}
\end{equation}
where $b_1\triangleq(A_2J_2-B_2K_2){\tau}_1{\bf \sigma}^2_{1,c}$, $b_2\triangleq(\alpha K_2-\beta J_2){\tau}_1{\bf \sigma}^2_{1,c}$ and $b_3\triangleq{\frac{(P_rE_2-L_2E_2-D_2K_2)B_2+(A_2D_2+C_2E_2)J_2}{J_2E_2}}$.

\item When the constraints \eqref{32b} and \eqref{32e} hold with equality, the optimal solution $\{P^*_1, P^*_2, \rho^* \}$ are given by\\
\begin{equation}\label{EQU-35} \vspace{-3pt}
\begin{split}
&P^*_1 = P_{max,1}, \quad P^*_2 = {\frac{\frac{{\tau}_1{\bf \sigma}^2_{1,c}}{1-\rho^*}+D_2}{E_2}},\quad\quad\\
&{\rho^*}=1-\sqrt{\frac{c_2-c_1}{E_2c_3}}.
\end{split}
\end{equation}
where $c_1\triangleq A_2{\tau}_1{\bf \sigma}^2_{1,c}$, $c_2\triangleq\beta{\tau}_1{\bf \sigma}^2_{1,c}$ and $c_3\triangleq{\frac{A_2D_2+B_2E_2 P_{max,1}+C_2E_2}{E_2}}$.

\item When the constraints \eqref{32c} and \eqref{32d} hold with equality, the optimal solution $\{P^*_1, P^*_2, \rho^* \}$ are given by\\
\begin{equation}\label{EQU-36} \vspace{-3pt}
\begin{split}
\hspace{0.5cm}&P^*_1= {\frac{\frac{{\tau}_2{\bf \sigma}^2_{2,c}}{1-\rho^*}+F_2}{G_2}},\hspace{0.3cm}P^*_2 = {\frac{P_r-L_2-P^*_1J_2}{K_2}},\\ &{\rho^*}=1-\sqrt{-\frac{d_1+d_2}{K_2G_2d_3}}.
\end{split}
\end{equation}
where $d_1\triangleq(B_2K_2-A_2J_2){\tau}_2{\bf \sigma}^2_{2,c}$, $d_2\triangleq(\beta J_2-\alpha K_2){\tau}_2{\bf \sigma}^2_{2,c}$ and $d_3\triangleq{\frac{(P_rG_2-L_2G_2-F_2J_2)A_2+(B_2F_2+C_2G_2)K_2}{G_2K_2}}$.

\item When the constraints \eqref{32c} and \eqref{32f} hold with equality, the optimal solution $\{P^*_1, P^*_2, \rho^* \}$ are given by\\
\begin{equation}\label{EQU-37} \vspace{-3pt}
\begin{split}
&P^*_1 = {\frac{\frac{{\tau}_2{\bf \sigma}^2_{2,c}}{1-\rho^*}+F_2}{G_2}}, \quad P^*_2 = P_{max,2},\quad\quad\\ &{\rho^*}=1-\sqrt{\frac{e_2-e_1}{G_2e_3}}.
\end{split}
\end{equation}
where $e_1\triangleq B_2{\tau}_2{\bf \sigma}^2_{2,c}$, $e_2\triangleq\alpha{\tau}_2{\bf \sigma}^2_{2,c}$ and $e_3\triangleq{\frac{B_2F_2+A_2G_2 P_{max,2}+C_2G_2}{G_2}}$.

\item When the two transmit power constraints \eqref{32d} and \eqref{32e} hold with equality, the optimal solution $\{P^*_1, P^*_2, \rho^* \}$ are given by\\
\begin{equation}\label{EQU-38} \vspace{-3pt}
\begin{split}
&P^*_1 = P_{max,1}, \quad P^*_2 = {\frac{P_r-L_2-J_2P_{max,1}}{K_2}},\\ &{\rho^*}=min\{1-\frac{{\tau}_1 {\bf \sigma}^2_{1,c}}{E_2P^*_2-D_2},1-\frac{{\tau}_2 {\bf \sigma}^2_{2,c}}{G_2P^*_1-F_2}\}.
\end{split}
\end{equation}

\item When the constraints \eqref{32d} and \eqref{32f} hold with equality, the optimal solution $\{P^*_1, P^*_2, \rho^* \}$ are given by\\
\begin{equation}\label{EQU-39} \vspace{-3pt}
\begin{split}
&P^*_1 = {\frac{P_r-L_2-K_2P_{max,2}}{J_2}}, \quad P^*_2 = P_{max,2},\\ &{\rho^*}=min\{1-\frac{{\tau}_1 {\bf \sigma}^2_{1,c}}{E_2P^*_2-D_2},1-\frac{{\tau}_2 {\bf \sigma}^2_{2,c}}{G_2P^*_1-F_2}\}.
\end{split}
\end{equation}

\item When the constraints \eqref{32e} and \eqref{32f} hold with equality, the optimal solution $\{P^*_1, P^*_2, \rho^* \}$ are given by\\
\begin{equation}\label{EQU-40} \vspace{-3pt}
\begin{split}
&P^*_1 = P_{max,1}, \quad P^*_2 = P_{max,2},\\ &{\rho^*}=min\{1-\frac{{\tau}_1 {\bf \sigma}^2_{1,c}}{E_2P^*_2-D_2},1-\frac{{\tau}_2 {\bf \sigma}^2_{2,c}}{G_2P^*_1-F_2}\}.
\end{split}
\end{equation}
\end{itemize}
{\proof} Please refer to Appendix B.

We compare all objective function values by substituting \eqref{EQU-33}$\sim$\eqref{EQU-40} into \eqref{32a} and select one $\{P^*_1, P^*_2, \rho^* \}$ as the optimal solution, if they lead to the greatest value of the objective function $f(\rho^*)$.

The proposed iterative algorithm to obtain the near optimal solution to problem \eqref{EQU-9} is summarized in the following Algorithm 1.

\vspace{-1pt}
\hrulefill
\par
{\footnotesize
\textbf{Algorithm 1} Finding a near optimal solution to problem \eqref{EQU-9}
\begin{itemize}
\item \textbf{Initialize} $P_1$, $P_2$ and ${\rho}$;
\item \textbf{Repeat}
\begin{itemize}
\item Update the beamforming matrixes ${\bf W}$ and ${\bf Q}_x$ for fixed $P_1$, $P_2$ and ${\rho}$ using the following steps: First, solve problem \eqref{EQU-30} by CVX, and obtain the optimal solution as $\tilde{\bf W}^*$ and ${\bf Q}_x$. Then, derive the optimal solution ${\bf W}$ of \eqref{EQU-25} by eigenvalue decomposition (EVD) of $\tilde{\bf W}^*$;
\item Update $P_1$, $P_2$ and ${\rho}$ with ${\bf W}$ and ${\bf Q}_x$ fixed using the following steps: First, check whether there exists some feasible solutions $\{P^*_1, P^*_2, \rho^* \}$ of problem (32). If yes, compute all the objective function values and select one $\{P^*_1, P^*_2, \rho^* \}$ corresponding to the largest value of objective function as the optimal solution to (32). Otherwise, exit the algorithm;
\end{itemize}
\item \textbf{Until} The difference between the value of objective function in \eqref{EQU-9} from one iteration to another is smaller than a pre-fixed predetermined threshold.
\end{itemize}}
\vspace{-9pt}
\hrulefill
\vspace{-0.1pt}

\subsection{Joint Beamforming and PS Optimization for DF-XOR relaying strategy} \vspace{-1pt}
In this subsection, we consider optimization problem \eqref{EQU-17} where the relay node adopts the DF-XOR two-way relaying strategy. Similar to problem \eqref{EQU-9}, we decouple problem \eqref{EQU-17} into two subproblems. It is worth noting that here different from \eqref{EQU-9}, two subproblems from \eqref{EQU-17} only involve the beamforming vectors and PS ratios, where ${P}_1$ and ${P}_2$ are not necessary to be optimized as they are determined via the constraints presented in \eqref{EQU-10}.

\textit{1) Optimize ${\bf Q}_s$ and ${\bf Q}_x$ for fixed ${\rho}$}: Note that when fixing ${\rho}$, the problem of optimizing variables ${\bf Q}_s$ and ${\bf Q}_x$ can be equivalent to
\begin{equation}\label{EQU-41} \vspace{-3pt}
\begin{split}
&\max_{{\bf Q}_s\succeq 0,{\bf Q}_x \succeq 0}~~\alpha({\bf g}^T_1 {\bf Q}_s {\bf g}^*_1+{\bf g}^T_1 {\bf Q}_x {\bf g}^*_1)\\
&\quad\quad\quad\quad\quad\quad+\beta({\bf g}^T_2 {\bf Q}_s {\bf g}^*_2+{\bf g}^T_2 {\bf Q}_x {\bf g}^*_2)\\
\textit{s.t.} ~~ &{\rm SINR}^{XOR}_i\geq {\tau}_i,~i=1,2.\\
&{\rm Tr}({\bf Q}_s)+{\rm Tr}({\bf Q}_x)\leq P_r,
\end{split}
\end{equation}
which is rewritten as
\begin{equation}\label{EQU-42} \vspace{-3pt}
\begin{split}
&\max_{{\bf Q}_s\succeq 0,{\bf Q}_x \succeq 0}~~{\rm Tr}({\bf A}_3{\bf Q}_s)+{\rm Tr}({\bf A}_3{\bf Q}_x)\\
\textit{s.t.} ~~ &{\rm Tr}({\bf B}_3{\bf Q}_s)-{\rm Tr}({\tau}_1{\bf B}_3{\bf Q}_x)\geq D_3,\\
&{\rm Tr}({\bf C}_3{\bf Q}_s)-{\rm Tr}({\tau}_2{\bf C}_3{\bf Q}_x)\geq E_3,\\
&{\rm Tr}({\bf Q}_s)+{\rm Tr}({\bf Q}_x)\leq P_r,
\end{split}
\end{equation}
where ${\bf A}_3\triangleq\alpha {\bf g}^*_1 {\bf g}^T_1 + \beta {\bf g}^*_2 {\bf g}^T_2$, ${\bf B}_3\triangleq{\bf g}^*_1 {\bf g}^T_1$, ${\bf C}_3\triangleq{\bf g}^*_2 {\bf g}^T_2$, $D_3\triangleq{\bf \sigma}^2_{1,d}+\frac{{\bf \sigma}^2_{1,c}}{1-\rho}{\tau}_1$ and $E_3\triangleq{\bf \sigma}^2_{2,d}+\frac{{\bf \sigma}^2_{2,c}}{1-\rho}{\tau}_2$.
It is easy to verify that \eqref{EQU-42} is a standard SDP problem. Thus, its optimal solution $\{{\bf Q}^*_s, {\bf Q}^*_x \}$ can be easily obtained using existing software, e.g., CVX \cite{Grant2010}.

\textit{2) Optimize ${\rho}$ for fixed ${\bf Q}_s$ and ${\bf Q}_x$}: In the second step, we need to optimize the PS ratio ${\rho}$ with the remaining variables fixed. The corresponding optimization problem can be formulated as
\begin{equation}\label{EQU-43} \vspace{-3pt}
\begin{split}
&\max_{\rho}~~{\frac{\alpha\eta T}{2}}{\rho}({\bf g}^T_1 {\bf Q}_s {\bf g}^*_1+{\bf g}^T_1 {\bf Q}_x {\bf g}^*_1)-{\frac{\alpha T}{2}}{P}_1\\
&\quad\quad\quad+{\frac{\beta\eta T}{2}}{\rho}({\bf g}^T_2 {\bf Q}_s {\bf g}^*_2+{\bf g}^T_2 {\bf Q}_x {\bf g}^*_2)-{\frac{\beta T}{2}}{P}_2\\
&\quad\textit{s.t.} ~~{\rm SINR}^{XOR}_i\geq {\tau}_i,~i=1,2.
\end{split}
\end{equation}
which is equivalent to
\begin{equation}\label{EQU-44} \vspace{-3pt}
\begin{split}
&\max_{\rho}~~(A_4+B_4){\rho}-{\frac{T}{2}}(\alpha{P}_1+\beta{P}_2)\\
\textit{s.t.} ~~ &C_4(1-\rho)\geq {\tau}_1{\bf \sigma}^2_{1,c}\\
&D_4(1-\rho)\geq {\tau}_2{\bf \sigma}^2_{2,c}
\end{split}
\end{equation}
where $A_4\triangleq{\frac{\alpha\eta T}{2}}({\bf g}^T_1 {\bf Q}_s {\bf g}^*_1+{\bf g}^T_1 {\bf Q}_x {\bf g}^*_1)$, $B_4\triangleq{\frac{\beta\eta T}{2}}({\bf g}^T_2 {\bf Q}_s {\bf g}^*_2+{\bf g}^T_2 {\bf Q}_x {\bf g}^*_2)$, $C_4\triangleq{\bf g}^T_1 {\bf Q}_s {\bf g}^*_1-({\bf g}^T_1 {\bf Q}_x {\bf g}^*_1+{\bf \sigma}^2_{1,d}){\tau}_1$ and $D_4\triangleq{\bf g}^T_2 {\bf Q}_s {\bf g}^*_2-({\bf g}^T_2 {\bf Q}_x {\bf g}^*_2+{\bf \sigma}^2_{2,d}){\tau}_2$.
According to the definition of ${P}_1$ and ${P}_2$ in \eqref{EQU-10}, the simplified PS design problem yields the following
\begin{equation}\label{EQU-45} \vspace{-3pt}
\begin{split}
&\max_{\rho}~~(A_4+B_4){\rho}\\
\textit{s.t.} ~~ &{\rho} \leq 1-\frac{{\tau}_1{\bf \sigma}^2_{1,c}}{C_4},\\
&{\rho} \leq 1-\frac{{\tau}_2{\bf \sigma}^2_{2,c}}{D_4}.
\end{split}
\end{equation}
It can be observed that the objective function in \eqref{EQU-45} achieves a higher value when one of the SINR constraints holds with equality. Hence, the optimal solution ${\rho^*}=min\{1-\frac{{\tau}_1{\bf \sigma}^2_{1,c}}{C_4},1-\frac{{\tau}_2{\bf \sigma}^2_{2,c}}{D_4}\}$ can be obtained from problem \eqref{EQU-45}.

To summarize, Algorithm 2 below summarizes the solution to \eqref{EQU-17}. Note that Algorithm 2 differs from Algorithm 1 in two main aspects: First, in step 1), the optimal beamforming matrixes $\{{\bf Q}^*_s, {\bf Q}^*_x \}$ can be easily obtained due to the absence of rank-one constraint; and second, step 2) only involves PS ratios ${\rho}$, ${P}_1$ and ${P}_2$ are not necessary to be optimized.

\vspace{-1pt}
\hrulefill
\par
{\footnotesize
\textbf{Algorithm 2} Finding the near optimal solution to problem \eqref{EQU-17}
\begin{itemize}
\item \textbf{Initialize} ${\rho}$;
\item \textbf{Repeat}
\begin{itemize}
\item solve problem \eqref{EQU-42} by CVX, and obtain the optimal solution of ${\bf Q}_s$ and ${\bf Q}_x $ in \eqref{EQU-41} which are denoted by ${\bf Q}^*_s$ and ${\bf Q}^*_x $, respectively;
\item Update PS ratios ${\rho}^*$ using \eqref{EQU-45} for fixed ${\bf Q}_s$ and ${\bf Q}_x$;
\end{itemize}
\item \textbf{Until} Termination criterion is satisfied.
\end{itemize}}
\vspace{-9pt}
\hrulefill
\vspace{-0.1pt}

\subsection{Joint Beamforming and PS Optimization for DF-SUP relaying strategy} \vspace{-1pt}
In this subsection, we consider that the DF-SUP relaying strategy is adopted at the relay node. To find the optimal solution of the joint optimization problem in \eqref{EQU-24}, We similarly decouple problem \eqref{EQU-24} into two subproblems, and then propose a two-tie iterative algorithm to obtain a near optimal solution of the original optimization problem.

\textit{1) Optimize ${\bf Q}_{s,1}$, ${\bf Q}_{s,2}$ and ${\bf Q}_x$ for fixed ${\rho}$}: In the first step, we need to optimize the beamforming matrices ${\bf Q}_{s,1}$, ${\bf Q}_{s,2}$ and ${\bf Q}_x$ with the PS ratio ${\rho}$ fixed. The corresponding optimization problem can be formulated as
\begin{equation}\label{EQU-46} \vspace{-3pt}
\begin{split}
&\max_{{\bf Q}_{s,1}\succeq 0,{\bf Q}_{s,2}\succeq 0,{\bf Q}_x \succeq 0}~~\alpha({\bf g}^T_1 {\bf Q}_{s,1} {\bf g}^*_1+{\bf g}^T_1 {\bf Q}_{s,2} {\bf g}^*_1\\
&\quad\quad\quad\quad\quad\quad\quad\quad\quad+{\bf g}^T_1 {\bf Q}_x {\bf g}^*_1)+\beta({\bf g}^T_2 {\bf Q}_{x,1} {\bf g}^*_2\\
&\quad\quad\quad\quad\quad\quad\quad\quad\quad+{\bf g}^T_2 {\bf Q}_{s,2} {\bf g}^*_2+{\bf g}^T_2 {\bf Q}_x {\bf g}^*_2)\\
&\textit{s.t.} ~~ {\rm SINR}^{SUP}_i\geq {\tau}_i,~i=1,2.\\
&\quad\quad{\rm Tr}({\bf Q}_{s,1})+{\rm Tr}({\bf Q}_{s,2})+{\rm Tr}({\bf Q}_x)\leq P_r,
\end{split}
\end{equation}
which is rewritten as
\begin{equation}\label{EQU-47} \vspace{-3pt}
\begin{split}
&\max_{{\bf Q}_{s,1}\succeq 0,{\bf Q}_{s,2}\succeq 0,{\bf Q}_x \succeq 0}~~{\rm Tr}({\bf A}_5({\bf Q}_{s,1}+{\bf Q}_{s,2}+{\bf Q}_x))\\
&\textit{s.t.} ~~ {\rm Tr}({\bf B}_5{\bf Q}_{s,2})-{\rm Tr}({\tau}_1{\bf B}_5{\bf Q}_x)\geq D_5,\\
&\quad\quad{\rm Tr}({\bf C}_5{\bf Q}_{s,1})-{\rm Tr}({\tau}_2{\bf C}_5{\bf Q}_x)\geq E_5,\\
&\quad\quad{\rm Tr}({\bf Q}_{s,1})+{\rm Tr}({\bf Q}_{s,2})+{\rm Tr}({\bf Q}_x)\leq P_r.
\end{split}
\end{equation}
where ${\bf A}_5\triangleq\alpha {\bf g}^*_1 {\bf g}^T_1 + \beta {\bf g}^*_2 {\bf g}^T_2$, ${\bf B}_5\triangleq{\bf g}^*_1 {\bf g}^T_1$, ${\bf C}_5\triangleq{\bf g}^*_2 {\bf g}^T_2$, $D_5\triangleq{\bf \sigma}^2_{1,d}+\frac{{\bf \sigma}^2_{1,c}}{1-\rho}{\tau}_1$ and $E_5\triangleq{\bf \sigma}^2_{2,d}+\frac{{\bf \sigma}^2_{2,c}}{1-\rho}{\tau}_2$.
Note that \eqref{EQU-47} is a standard SDP problem. Thus, its optimal solution can be easily obtained \cite{Grant2010}.

\textit{2) Optimize ${\rho}$ for fixed ${\bf Q}_{s,1}$, ${\bf Q}_{s,2}$ and ${\bf Q}_x$}: In the second step, we need to optimize the PS ratio ${\rho}$ with the remaining variables fixed. The corresponding optimization problem can be formulated as
\begin{equation}\label{EQU-48} \vspace{-3pt}
\begin{split}
&\max_{\rho}~~{\frac{\alpha\eta T}{2}}{\rho}({\bf g}^T_1 {\bf Q}_{s,1} {\bf g}^*_1+{\bf g}^T_1 {\bf Q}_{s,2} {\bf g}^*_1\\
&\quad\quad\quad+{\bf g}^T_1 {\bf Q}_x {\bf g}^*_1)-{\frac{\alpha T}{2}}{P}_1+{\frac{\beta\eta T}{2}}{\rho}({\bf g}^T_2 {\bf Q}_{s,1} {\bf g}^*_2\\
&\quad\quad\quad+{\bf g}^T_2 {\bf Q}_{s,2} {\bf g}^*_2+{\bf g}^T_2 {\bf Q}_x {\bf g}^*_2)-{\frac{\beta T}{2}}{P}_2\\
&\quad\textit{s.t.} ~~{\rm SINR}^{SUP}_i\geq {\tau}_i,~i=1,2.
\end{split}
\end{equation}
which is equivalent to
\begin{equation}\label{EQU-49} \vspace{-3pt}
\begin{split}
&\max_{\rho}~~(A_6+B_6){\rho}-{\frac{T}{2}}(\alpha{P}_1+\beta{P}_2)\\
\textit{s.t.} ~~ &C_6(1-\rho)\geq {\tau}_1{\bf \sigma}^2_{1,c},\\
&D_6(1-\rho)\geq {\tau}_2{\bf \sigma}^2_{2,c},
\end{split}
\end{equation}
where $A_6\triangleq{\frac{\alpha\eta T}{2}}({\bf g}^T_1 {\bf Q}_{s,1} {\bf g}^*_1+{\bf g}^T_1 {\bf Q}_{s,2} {\bf g}^*_1+{\bf g}^T_1 {\bf Q}_x {\bf g}^*_1)$, $B_6\triangleq{\frac{\beta\eta T}{2}}({\bf g}^T_2 {\bf Q}_{s,1} {\bf g}^*_2+{\bf g}^T_2 {\bf Q}_{s,2} {\bf g}^*_2+{\bf g}^T_2 {\bf Q}_x {\bf g}^*_2)$, $C_6\triangleq{\bf g}^T_1 {\bf Q}_{s,2} {\bf g}^*_1-({\bf g}^T_1 {\bf Q}_x {\bf g}^*_1+{\bf \sigma}^2_{1,d}){\tau}_1$ and $D_6\triangleq{\bf g}^T_2 {\bf Q}_{s,1} {\bf g}^*_2-({\bf g}^T_2 {\bf Q}_x {\bf g}^*_2+{\bf \sigma}^2_{2,d}){\tau}_2$. Since ${P}_1$ and ${P}_2$ are determined based on the first phase, problem \eqref{EQU-49} is simplified as
\begin{equation}\label{EQU-50} \vspace{-3pt}
\begin{split}
&\max_{\rho}~~(A_6+B_6){\rho}\\
\textit{s.t.} ~~ &{\rho} \leq 1-\frac{{\tau}_1{\bf \sigma}^2_{1,c}}{C_6},\\
&{\rho} \leq 1-\frac{{\tau}_2{\bf \sigma}^2_{2,c}}{D_6}.
\end{split}
\end{equation}
Similar to the problem \eqref{EQU-45}, the optimal PS solution ${\rho^*}=min\{1-\frac{{\tau}_1{\bf \sigma}^2_{1,c}}{C_6},1-\frac{{\tau}_2{\bf \sigma}^2_{2,c}}{D_6}\}$ can be obtained from problem \eqref{EQU-50}.

Similar to Algorithm 2, the proposed iterative algorithm to problem \eqref{EQU-24} is summarized in Algorithm 3.

\vspace{-1pt}
\hrulefill
\par
{\footnotesize
\textbf{Algorithm 3} Finding the near optimal solution to problem \eqref{EQU-24}
\begin{itemize}
\item \textbf{Initialize} ${\rho}$;
\item \textbf{Repeat}
\begin{itemize}
\item Update the beamforming matrixes ${\bf Q}_{s,1}$, ${\bf Q}_{s,2}$ and ${\bf Q}_x$ with \eqref{EQU-47} for fixed ${\rho}$;
\item Update PS ratios ${\rho}$ using \eqref{EQU-50} for fixed ${\bf Q}_{s,1}$, ${\bf Q}_{s,2}$ and ${\bf Q}_x$;
\end{itemize}
\item \textbf{Until} Termination criterion is satisfied.
\end{itemize}}
\vspace{-9pt}
\hrulefill
\vspace{-0.1pt}

\section{Simulation results} \vspace{-1pt}
In this section, we numerically evaluate the performance of the proposed energy harvesting schemes. The channel vector ${\bf h}_i$ and ${\bf g}_i$ are set to be Rayleigh fading, i.e., the elements of each channel matrix or vector are complex Gaussian random variables with zero mean and unit variance. The channel gain is modeled by the distance path loss model\cite{Wang2013}, given as $g_{i,j}=c \cdot d^{-n}_{i,j}$, where $c$ is an attenuation constant set as 1, $n$ is the path loss exponent and fixed at $3$, and $d_{i,j}$ denotes the distance between nodes $i$ and $j$. For simplicity, we assume that the noise power at all the destinations are the same, i.e., ${\bf \sigma}^2_{i,c}={\bf \sigma}^2_{i,d}={\bf \sigma}^2_r={\bf \sigma}^2=0$ ${\rm dBm}$, ${\forall i}$, and the energy conversion efficiency $\eta=50\%$. Moreover, the maximum transmit powers at the two sources, if not specified, are set as ${P}_{max,1}={P}_{max,2}={P}_{max}=5$ ${\rm dBm}$. In all simulations, the weighted sum power of the relay network is computed by using 1000 randomly generated channel realizations.

\begin{figure}[t]
  \centering
  \subfigure[]{
    \label{AF1} 
    \includegraphics[scale=0.605]{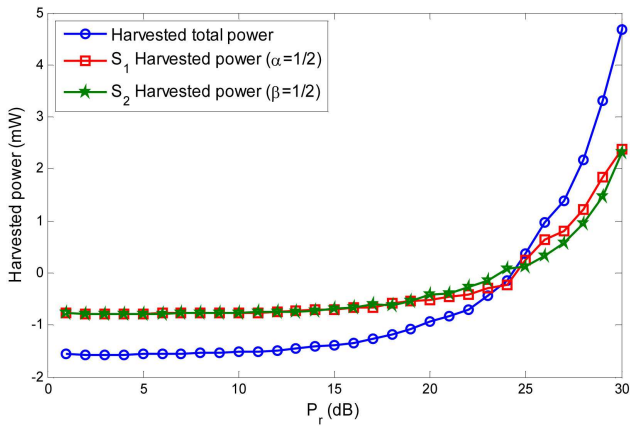}}
  \hspace{0in}
  \subfigure[]{
    \label{AF2} 
    \includegraphics[scale=0.605]{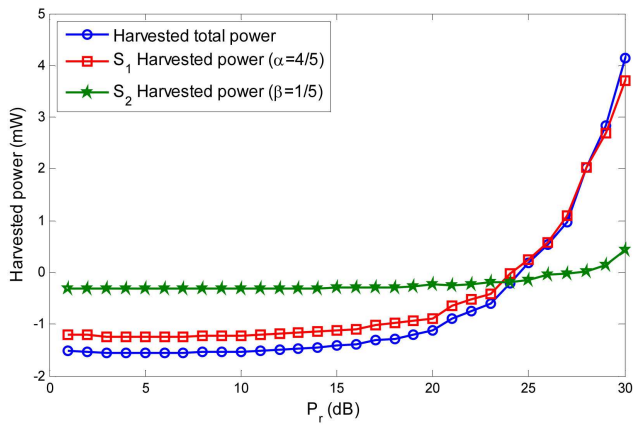}}
  \caption{Performance comparison for AF Relaying Strategy when changing $P_r$. (a) Harvested power of the two source nodes with same priority. (b) Harvested power of the two source nodes with different priority.}
  \label{AF} 
  \vspace{-0.7cm}
\end{figure}

\begin{figure}[t]
  \centering
  \subfigure[]{
    \label{XOR1} 
    \includegraphics[scale=0.62]{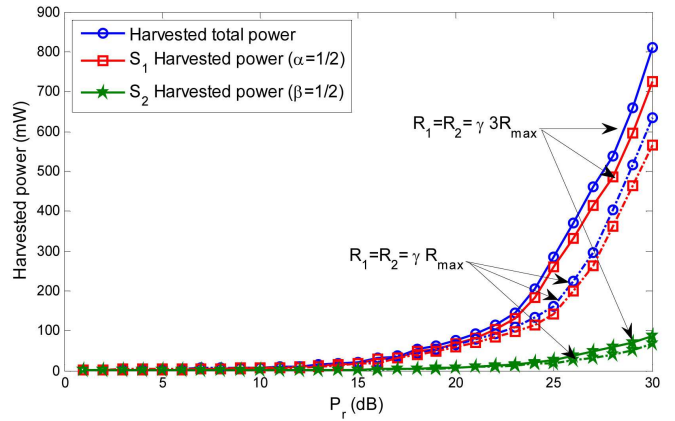}}
  \hspace{0in}
  \subfigure[]{
    \label{XOR2} 
    \includegraphics[scale=0.62]{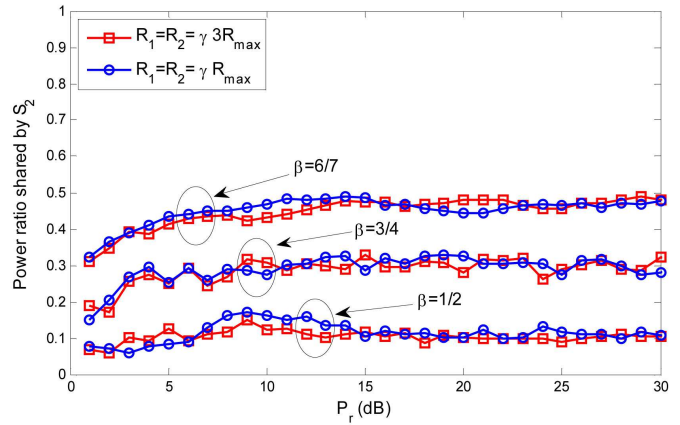}}
  \caption{Performance comparison for DF-XOR Relaying Strategy at $\gamma=0.1$ when changing $P_r$. (a) Harvested power of the two source nodes with same priority. (b) Power ratio shared by $S_2$ with different priority.}
  \label{XOR} 
  \vspace{-0.5cm}
\end{figure}

\begin{figure}[t]
  \centering
  \subfigure[]{
    \label{SUP1} 
    \includegraphics[scale=0.62]{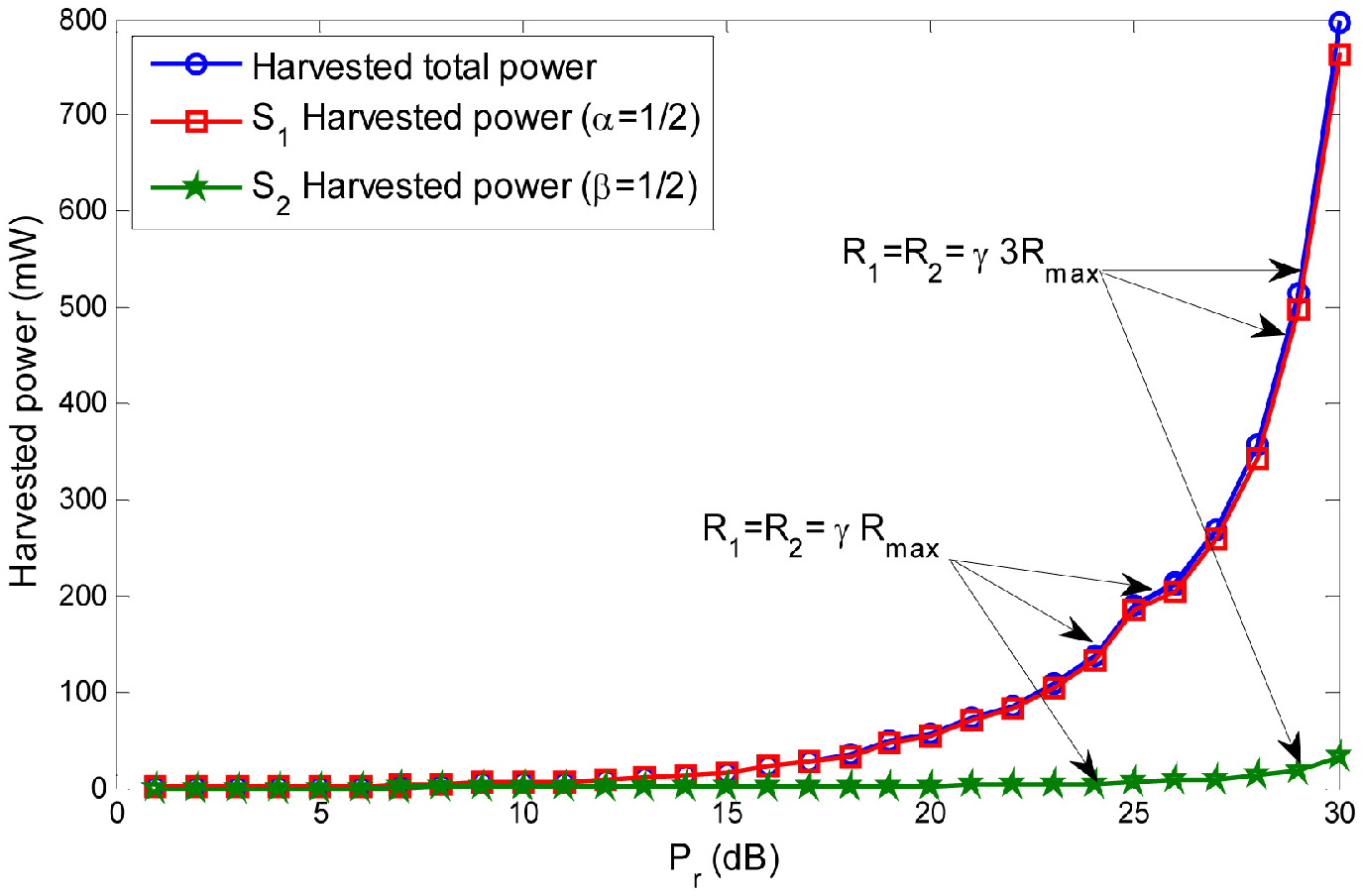}}
  \hspace{0in}
  \subfigure[]{
    \label{SUP2} 
    \includegraphics[scale=0.62]{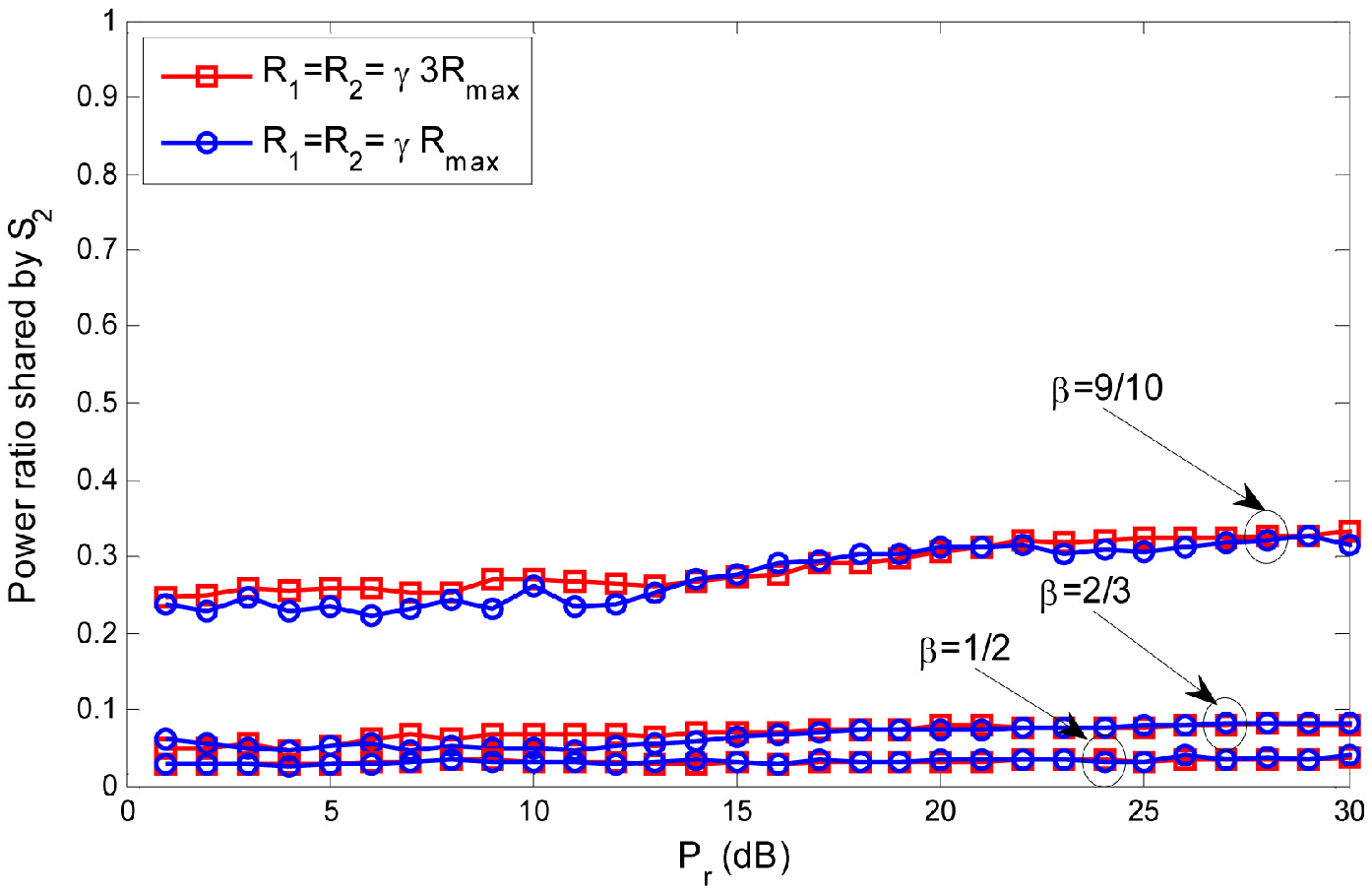}}
  \caption{Performance comparison for DF-SUP Relaying Strategy at $\gamma=0.1$ when changing $P_r$. (a) Harvested power of the two source nodes with same priority. (b) Power ratio shared by $S_2$ with different priority.}
  \label{SUP} 
  \vspace{-0.3cm}
\end{figure}

In Figs.~\ref{AF}-\ref{SUP}, we present the weighted sum power harvested for different relaying strategies at different distance of two sources when the relay node is equipped with $N=4$ transmit antennas. In Figs.~\ref{AF}, we illustrate the harvested power of the two source nodes when the relay node adopts AF relaying strategy. Specifically, the distances of the two source nodes are symmetric, i.e., $d_{R,S_1}=d_{R,S_2}=1$ meter. For comparison, two different energy weights with $\alpha=\beta=1/2$ and $\alpha=4/5$ and $\beta=1/5$ are simulated for this scenario. From simulation results, in Fig.~\ref{AF1}, we find that when $S_1$ and $S_2$ have the same priority, the two nodes can achieve a fair energy efficiency. However, in Fig.~\ref{AF2}, when $S_1$ and $S_2$ have different priorities, node $S_1$ can harvest more energy since its energy weight factor is set to a larger value. Moreover, note that when the relay transmit power, $P_r$, is low, the harvested energy at the nodes is negative, which implies that the harvested power at nodes from the relay is smaller than the consumed power for signal transmission.

In Figs.~\ref{XOR} and~\ref{SUP}, we illustrate the harvested power at the two source nodes, i.e., $P_{S_1}$ and $P_{S_2}$, and the shared power ratio at the $S_2$ node, i.e., $\frac{P_{S_2}}{P_{S_1}+P_{S_2}}$, in different rate requirements $R_1=R_2= \gamma R_{max}$ and $R_1=R_2= \gamma 3R_{max}$, where $R_{max}=\frac{1}{2}{\rm log_2}(1+\frac{P_{max}}{{\bf \sigma}^2})$. It is noted that in asymmetric scenario, i.e., $d_{R,S_1}=1$ meter and $d_{R,S_2}=2$ meters, in Fig.~\ref{XOR1} and with $d_{R,S_1}=1$ meter and $d_{R,S_2}=3$ meters in Fig.~\ref{SUP1}, although two source nodes $S_1$ and $S_2$ have same priority, the node $S_2$ still harvests much lower energy in different rate requirements. The main reason is that the location of $S_2$ is far away from the relay node $R$, which could result in very small channel gain as compared to the near node. This coupled effect is referred to as the doubly-near-far problem \cite{Bi2015}. However, when with higher priority, i.e., $\beta=3/4$ and $\beta=6/7$, in Fig.~\ref{XOR2} and with $\beta=2/3$ and $\beta=9/10$ in Fig.~\ref{SUP2}, we find that node $S_2$ all can share more power for the harvested total power in different rate requirements. This indicates that under the asymmetric scenario, the far node will be able to harvest more energy when its energy weight factor is set to a larger value, which can provide an effective solution to the doubly-near-far problem. In addition, note that here different the results in Figs.~\ref{AF1} and~\ref{AF2}, the harvested energy at the nodes is positive, which implies that the harvested power at nodes from the relay is greater than the consumed power for signal transmission.

\begin{figure}[t]
\begin{centering}
\includegraphics[scale=0.62]{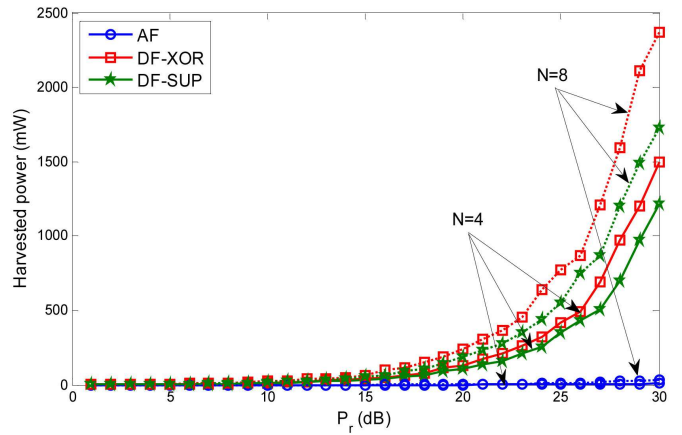}
\vspace{-0.5cm}
\caption{Performance comparison for different relay strategies with different number of antennas at relay when changing $P_r$.} \label{compare}
\end{centering}
\vspace{-0.2cm}
\end{figure}

 In Fig.~\ref{compare}, we illustrate the weighted sum power harvested for different relay strategies with different number of antennas at relay when changing $P_r$. For fair comparison, the priorities, the rate requirements and the distances of two source nodes are set to be the same, i.e., $\alpha=\beta=1/2$, $R_1=R_2= 0.1 R_{max}$ and $d_{R,S_1}=d_{R,S_2}=1$ meter, and simulated for each scenario. From simulation results, we find that the DF-XOR relaying strategy achieves the best performance, and the DF-SUP relaying strategy outperforms the AF relaying strategy. This indicates that DF relaying strategy has a higher EH efficiency due to the assumption that the relay has enough processing ability to correctly decode the received signals. Moreover, combining the information using XOR is better than using superposition since the power of the relay node can be used more efficiently in the DF-XOR relaying strategy. In addition, it is observed that when the number of transmit antennas increases $(N=4\rightarrow8)$, the all three considered two-way relaying strategies achieve better performance. This demonstrates the significant benefit by applying large or even massive antenna arrays for efficiently implementing TWR SWIPT systems in practice.

\begin{figure}[t]
\begin{centering}
\includegraphics[scale=0.62]{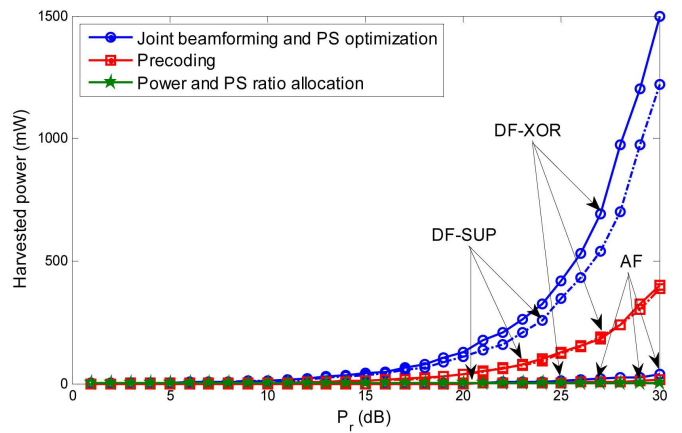}
\vspace{-0.5cm}
\caption{Performance comparison with different schemes at $\alpha=\beta=1/2$, $R_1=R_2= 0.1 R_{max}$ and $d_{R,S_1}=d_{R,S_2}=1$.} \label{compare1}
\end{centering}
\vspace{-0.2cm}
\end{figure}

Finally, in Fig.~\ref{compare1}, we compare the proposed joint beamforming and PS optimization scheme with the other two schemes, i.e., only precoding without power and PS ratio allocation, and only power and PS ratio allocation without precoding scheme. For fairness, the setting of each node is the same with the one in Fig.~\ref{compare}. From simulation results, for three considered two-way relaying strategies, we find that the joint beamforming and PS optimization scheme all achieves the best performance as it uses the degrees of the freedom of both power, PS ratio allocation and precoding. Moreover, Fig.~\ref{compare1} also shows that the precoding only scheme can improves the system performance and it performs much better than the power and PS ratio allocation only scheme.

\section{Conclusions}
This paper has studied the joint energy transmit beamforming and power splitting design for a multi-antenna TWR system based simultaneous wireless information and power transfer (SWIPT). The weighted sum power harvested at all source nodes was maximized subject to given SINR constraints at different source nodes and transmitted power constraints at relay node. Considering three different relaying strategies, the design problems are formulated as three nonconvex quadratically constrained problem, which are decoupled into two subproblems that can be solved separately by applying suitable optimization tools. The performance of three relay strategies were compared and some practical implementation issues were discussed. Simulation results showed when the priority and the distance of two source nodes are symmetric, the DF-XOR relaying strategy performs better than the other two strategies. While the distances of two source nodes are asymmetric, for all three considered relaying strategies, the furthest node can harvests more energy when its energy weight factor is set to a larger value.


\appendices
\section{Proof of Lemma 1}
Note that, Problem \eqref{EQU-30} is a quasi-convex optimization problem, its optimal solution can be easily obtained \cite{Grant2010}. Let us denote the optimal solution of
$\tilde{\bf W}$ and ${\bf Q}_x$ in \eqref{EQU-30} by $\tilde{\bf W}^*$ and ${\bf Q}^*_x$, respectively. It is easy to verify that $\tilde{\bf W}^*$ is an optimal solution of
the following optimization problem
\begin{equation}\label{EQU-51} \vspace{-3pt}
\begin{split}
&\max_{{\tilde{\bf W}}\succeq 0}~~{\rm Tr}({\bf A}_1 {\tilde{\bf W}})\\
\textit{s.t.} ~~&{\rm Tr}({\bf C}^i_1 {\tilde{\bf W}})\geq D^i_1+{\rm Tr}({\tau}_i{\bf g}^*_i {\bf g}^T_i {\bf Q}^*_x),~i=1,2.\\
&{\rm Tr}({\bf E}_1{\tilde{\bf W}})\leq P_r-{\rm Tr}({\bf Q}^*_x),\\
\end{split}
\end{equation}
According to Lemma 3.1 in \cite{Huang2010}, there exists an optimal solution $\tilde{\bf W}^*$ for the problem \eqref{EQU-51} such that
\begin{equation}\label{EQU-52} \vspace{-3pt}
({\rm Rank}(\tilde{\bf W}^*))^2\leq 3
\end{equation}
It can be verified that $\tilde{\bf W}^* \neq {\bf 0}$. Thus, from \eqref{EQU-52}, we have ${\rm Rank}(\tilde{\bf W}^*)=1$. The proof of Lemma 1 is thus completed.\\

\section{Proof of Lemma 2}
Suppose problem (32) is feasible and let $\{P^*_1, P^*_2, \rho^* \}$ and $f(\cdot)$ denote the optimal solution and the objective function, respectively. Next, we show that for problem (32), with the optimal solution $\{P^*_1, P^*_2, \rho^* \}$, either the SINR constraints or the transmit power constraints must hold with equality. Now we prove this result by contradiction. Namely, if the above conditions are not satisfied, we can find another solution of (32), denoted by $\{P^{\dagger}_1, P^{\dagger}_2, \rho^{\dagger} \}$, which achieves a higher value for the objective function $f(P^{\dagger}_1, P^{\dagger}_2, \rho^{\dagger})$ than $f(P^*_1, P^*_2, \rho^*)$. First, suppose that the two SINR constraints do not hold with equality. In this case, note that in problem (32), the two SINR constraints are equivalent to
\begin{equation}\label{EQU-53} \vspace{-3pt}
{\rho}^{\dagger} \leq 1-{\frac{{\tau}_1{\bf \sigma}^2_{1,c}}{EP_2-D}},
\end{equation}
and
\begin{equation}\label{EQU-54} \vspace{-3pt}
{\rho}^{\dagger} \leq 1-{\frac{{\tau}_2{\bf \sigma}^2_{2,c}}{GP_1-F}}.
\end{equation}
It can be observed that when PS solution $\rho^{\dagger}=min\{1-{\frac{{\tau}_1{\bf \sigma}^2_{1,c}}{EP_2-D}},1-{\frac{{\tau}_2{\bf \sigma}^2_{2,c}}{GP_1-F}}\}$, i.e., one of the SINR constraints holds with equality, the objective function $f(P^{\dagger}_1, P^{\dagger}_2, \rho^{\dagger})=(A_2 \rho^{\dagger}-\beta)P^{\dagger}_2+(B_2 \rho^{\dagger}-\alpha)P^{\dagger}_1+C_2 \rho^{\dagger}$ achieves a higher value as the same transmit power constraints. Hence, this assumption is not true. Next, consider the case that the transmit power constraints do not hold with equality. In this case, if we want the PS solution $\rho$ to increase to achieve a higher value of objective function, the transmit power solution $P_i$ increases. Then, for the three transmit power constraints \eqref{32d}, \eqref{32e} and \eqref{32f}, there exists at least a power constraint that holds with equality. Thus, the assumption that the transmit power constraints do not hold with equality, also cannot be true. To summarize, for the optimal solution $\{P^*_1, P^*_2, \rho^* \}$, either the SINR or transmit power constraint must hold with equality.

Moreover, given the optimal solution $\{P^*_1, P^*_2, \rho^* \}$, we can prove that two constraints of problem (32) are achieved with equality. First, suppose that the optimal solution $\{P^*_1, P^*_2, \rho^* \}$ can be obtained if and only if the SINR constraint \eqref{32b} holds with equality. In this case, we can easily find another solution of $P_1$ for (32) while two transmit power constraints \eqref{32d} or \eqref{32e} hold with equality. We denote this solution of $P_1$ by $\tilde{P}^*_1$. We can verify that this solution achieves a higher value for the objective function $f(\tilde{P}^*_1, P^*_2, \rho^*)$ than $f(P^*_1, P^*_2, \rho^*)$. Hence, this assumption cannot be true. Similarly, for all the other assumptions where if and only if a constraint holds with equality, we can easily prove these assumptions cannot be true too. In conclusion, for the optimal solution $\{P^*_1, P^*_2, \rho^* \}$, there exist at least two constraints of problem (32) holding with equality. Based on the above observations, we can separate the problem (32) into ten cases by setting the two SINR constraints \eqref{32b}, \eqref{32c} and the three transmit power constraints \eqref{32d}, \eqref{32e}, \eqref{32f} hold with equality. However, in the ten cases, when the constraints \eqref{32b} and \eqref{32f} hold with equality, if we want the objective function $f(\cdot)$ to increase, the transmit power solution $P_1$ increases, which leads to the conclusion that the constraints \eqref{32d} or \eqref{32e} are active. Hence, this combination is included in \eqref{32d} and \eqref{32f} or \eqref{32e} and \eqref{32f} so that this case can be removed. Similarly, constraints \eqref{32c} and \eqref{32e} combination will also be removed since this case was contained in \eqref{32d} and \eqref{32e} or \eqref{32e} and \eqref{32f}. Therefore, the optimal solution $\{P^*_1, P^*_2, \rho^* \}$ of problem (32) is able to be obtained in closed-form by comparing following eight cases.

When the two SINR constraints \eqref{32b} and \eqref{32c} hold with equality, we obtain the following two equations
\begin{equation}\label{EQU-55} \vspace{-3pt}
\begin{split}
P^*_2 = {\frac{\frac{{\tau}_1{\bf \sigma}^2_{1,c}}{1-\rho^*}+D_2}{E_2}}, \quad P^*_1 = {\frac{\frac{{\tau}_2{\bf \sigma}^2_{2,c}}{1-\rho^*}+F_2}{G_2}},
\end{split}
\end{equation}
By substituting \eqref{EQU-55} into \eqref{32a}, the objective function $f(\rho^*)$ can be equivalently written as
\begin{equation}\label{EQU-56} \vspace{-3pt}
\begin{split}
f(\rho^*)=&A_2{\rho^*}{\frac{\frac{{\tau}_1{\bf \sigma}^2_{1,c}}{1-\rho^*}+D_2}{E_2}}+
B_2\rho^*{\frac{\frac{{\tau}_2{\bf \sigma}^2_{2,c}}{1-\rho^*}+F_2}{G_2}}\\
&-\alpha{\frac{\frac{{\tau}_2{\bf \sigma}^2_{2,c}}{1-\rho^*}+F_2}{G_2}}-\beta{\frac{\frac{{\tau}_1{\bf \sigma}^2_{1,c}}{1-\rho^*}+D_2}{E_2}}+C_2\rho^*,
\end{split}
\end{equation}
which is further equivalent to
\begin{equation}\label{EQU-57} \vspace{-3pt}
\begin{split}
f(\rho^*)={\frac{a_1(\rho^*)^2+a_2\rho^*-a_3}{E_2G_2(1-\rho^*)}-a_4}.
\end{split}
\end{equation}
where $a_1$, $a_2$, $a_3$ are defined as in \eqref{EQU-33} and $a_4\triangleq\alpha F_2/G_2+\beta D_2/E_2$.
Hence, problem (32) is simplified as
\begin{equation}\label{EQU-58} \vspace{-3pt}
\begin{split}
&\max_{\rho^*} ~~ f(\rho^*)\\
\textit{s.t.} ~~&{P^*_1 J_2+P^*_2 K_2}\leq P_r-L_2,\\
&0<P^*_1\leq {P}_{max,1},\\
&0<P^*_2\leq {P}_{max,2},\\
&{0<\rho^*<1}.
\end{split}
\end{equation}
To proceed to solve \eqref{EQU-58}, we have the following lemma.

\textit{Lemma 3}\label{lemma3}: The optimal solution ${\rho^*}=1-\sqrt{\frac{a_1+a_2-a_3}{a_1}}$ can be obtained from problem \eqref{EQU-58} while $0<a_2-a_3<-a_1$.
{\proof} Please refer to Appendix C.\\
Then we obtain the optimal solution in \eqref{EQU-33}.

When the constraints \eqref{32b} and \eqref{32d} hold with equality, we obtain the following two equations
\begin{equation}\label{EQU-59} \vspace{-3pt}
\begin{split}
P^*_2 = {\frac{\frac{{\tau}_1{\bf \sigma}^2_{1,c}}{1-\rho^*}+D_2}{E_2}}, \quad P^*_1  = {\frac{P_r-L_2-P^*_2K_2}{J_2}},
\end{split}
\end{equation}
By substituting \eqref{EQU-59} into \eqref{32a}, the objective function $f(\rho^*)$ can be equivalently written as
\begin{equation}\label{EQU-60} \vspace{-3pt}
\begin{split}
f(\rho^*)={\frac{-b_3J_2E_2{(\rho^*)}^2+(b_3J_2E_2+b_1)\rho^*+b_2}{J_2E_2(1-\rho^*)}-b_4},
\end{split}
\end{equation}
where $b_1$, $b_2$, $b_3$ are defined as in \eqref{EQU-34} and $b_4\triangleq{\frac{(P_rE_2-L_2E_2-D_2K_2)\alpha+D_2J_2\beta}{J_2E_2}}$.
\eqref{EQU-60} is equivalent to
\begin{equation}\label{EQU-61} \vspace{-3pt}
\begin{split}
f(\rho^*)={-b_3(1-\rho^*)+\frac{b_1+b_2}{J_2E_2(1-\rho^*)}-\frac{b_1}{J_2E_2}+b_3-b_4}.
\end{split}
\end{equation}
Then problem (32) is equivalent to the following problem
\begin{equation}\label{EQU-62} \vspace{-3pt}
\begin{split}
&\max_{\rho^*} ~~ f(\rho^*)\\
\textit{s.t.} ~~&{\frac{\frac{{\tau}_2{\bf \sigma}^2_{2,c}}{1-\rho^*}+F_2}{G_2}}\leq P^*_1\leq P_{max,1},\\
&0<P^*_2\leq P_{max,2},\\
&{0<\rho^*<1}.
\end{split}
\end{equation}
Similar to Lemma 3, when $b_3>0$ and $b_1+b_2<0$, the objective function $f(\rho^*)$ must have a maximum value, which can be further derived from ${-b_3(1-{\rho^*})}={\frac{b_1+b_2}{J_2E_2(1-\rho^*)}}$. On the other hand, to guarantee the optimal solution ${\rho^*}$ satisfying $0<{\rho^*}<1$, we have $b_1+b_2+J_2E_2b3>0$, which results in an optimal solution ${\rho^*}$ of problem \eqref{EQU-62} given as
\begin{equation}\label{EQU-63} \vspace{-3pt}
\begin{split}
{\rho^*}=1-\sqrt{-\frac{b_1+b_2}{J_2E_2b_3}}.
\end{split}
\end{equation}
we thus obtain the solution given in \eqref{EQU-34}.

When the constraints \eqref{32b} and \eqref{32e} hold with equality, we obtain the following two equations
\begin{equation}\label{EQU-64} \vspace{-3pt}
\begin{split}
P^*_2 = {\frac{\frac{{\tau}_1{\bf \sigma}^2_{1,c}}{1-\rho^*}+D_2}{E_2}}, \quad P^*_1 = P_{max,1},
\end{split}
\end{equation}
By substituting \eqref{EQU-64} into \eqref{32a}, the objective function $f(\rho^*)$ can be written as
\begin{equation}\label{EQU-65} \vspace{-3pt}
\begin{split}
f(\rho^*)={\frac{-c_3E_2{(\rho^*)}^2+(c_3E_2+c_1)\rho^*-c_2}{E_2(1-\rho^*)}-c_4},
\end{split}
\end{equation}
where $c_1$, $c_2$, $c_3$ are defined as in \eqref{EQU-35} and $c_4\triangleq{\frac{\alpha E_2 P_{max,1}+\beta D_2}{E_2}}$.
\eqref{EQU-65} is equivalent to
\begin{equation}\label{EQU-66} \vspace{-3pt}
\begin{split}
f(\rho^*)={-c_3(1-\rho^*)+\frac{c_1-c_2}{E_2(1-\rho^*)}-\frac{c_1}{E_2}+c_3-c_4}.
\end{split}
\end{equation}
Then problem (32) is simplified as
\begin{equation}\label{EQU-67} \vspace{-3pt}
\begin{split}
&\max_{\rho^*} ~~ f(\rho^*)\\
\textit{s.t.} ~~&P^*_1 \geq {\frac{\frac{{\tau}_2{\bf \sigma}^2_{2,c}}{1-\rho^*}+F_2}{G_2}},\\
&{P^*_1 J_2+P^*_2 K_2}\leq P_r-L_2,\\
&0<P^*_2\leq P_{max,2},\\
&{0<\rho^*<1}.
\end{split}
\end{equation}
Due to the fact $c_3>0$, i.e., $-c_3<0$. Similar to Lemma 3, if $\frac{c_1-c_2}{E_2}<0$, i.e., $c_2>c_1$, the objective function $f(\rho^*)$ must have a maximum value, which can be inferred from a fact that ${-c_3(1-{\rho^*})}={\frac{c_1-c_2}{E_2(1-\rho^*)}}$. Note that, to guarantee the optimal solution ${\rho^*}$ satisfying $0<{\rho^*}<1$, $c_2$ must satisfy $c_2<c_1+E_2c_3$. As a result, the optimal solution ${\rho^*}$ of problem \eqref{EQU-67} can be derived as
\begin{equation}\label{EQU-68} \vspace{-3pt}
\begin{split}
{\rho^*}=1-\sqrt{\frac{c_2-c_1}{E_2c_3}}.
\end{split}
\end{equation}
Then we obtain the optimal solution in \eqref{EQU-35}.

When the constraints \eqref{32c} and \eqref{32d} hold with equality, we obtain the following two equations
\begin{equation}\label{EQU-69} \vspace{-3pt}
P^*_1= {\frac{\frac{{\tau}_2{\bf \sigma}^2_{2,c}}{1-\rho^*}+F_2}{G_2}},\quad P^*_2 = {\frac{P_r-L_2-P^*_1J_2}{K_2}},
\end{equation}
By substituting \eqref{EQU-69} into \eqref{32a}, the objective function $f(\rho^*)$ can be equivalently written as
\begin{equation}\label{EQU-70} \vspace{-3pt}
\begin{split}
f(\rho^*)={\frac{-d_3K_2G_2({\rho^*})^2+(d_3K_2G_2+d_1)\rho^*+d_2}{K_2G_2(1-\rho^*)}-d_4},
\end{split}
\end{equation}
where $d_1$, $d_2$, $d_3$ are defined as in \eqref{EQU-36} and $d_4\triangleq{\frac{(P_rG_2-L_2G_2-F_2J_2)\beta+F_2K_2\alpha}{G_2K_2}}$.
\eqref{EQU-70} is equivalent to
\begin{equation}\label{EQU-71} \vspace{-3pt}
\begin{split}
f(\rho^*)={-d_3(1-\rho^*)+\frac{d_1+d_2}{K_2G_2(1-\rho^*)}-\frac{d_1}{K_2G_2}+d_3-d_4}.
\end{split}
\end{equation}
Then problem (32) can be rewritten as
\begin{equation}\label{EQU-72} \vspace{-3pt}
\begin{split}
&\max_{\rho^*} ~~ f(\rho^*)\\
\textit{s.t.} ~~&{\frac{\frac{{\tau}_1{\bf \sigma}^2_{1,c}}{1-\rho^*}+D_2}{E_2}}\leq P^*_2\leq P_{max,2},\\
&0<P^*_1\leq P_{max,1},\\
&{0<\rho^*<1}.
\end{split}
\end{equation}
Similar to Lemma 3, when $d_3>0$ and $d_1+d_2<0$, the objective function $f(\rho^*)$ must have a maximum value, which can be further derived from ${-d_3(1-{\rho^*})}={\frac{d_1+d_2}{K_2G_2(1-\rho^*)}}$. When $d_1+d_2+K_2G_2d_3>0$, the optimal solution ${\rho^*}$ of problem \eqref{EQU-72} can be derived as
\begin{equation}\label{EQU-73} \vspace{-3pt}
\begin{split}
{\rho^*}=1-\sqrt{-\frac{d_1+d_2}{K_2G_2d_3}}.
\end{split}
\end{equation}
Then we obtain the optimal solution in \eqref{EQU-36}.

When the constraints \eqref{32c} and \eqref{32f} hold with equality, we obtain the following two equations
\begin{equation}\label{EQU-74} \vspace{-3pt}
P^*_1 = {\frac{\frac{{\tau}_2{\bf \sigma}^2_{2,c}}{1-\rho^*}+F_2}{G_2}},\quad P^*_2 = P_{max,2},
\end{equation}
By substituting \eqref{EQU-74} into \eqref{32a}, the objective function $f(\rho^*)$ can be equivalently written as
\begin{equation}\label{EQU-75} \vspace{-3pt}
\begin{split}
f(\rho^*)={\frac{-e_3G_2({\rho^*})^2+(e_3G_2+e_1)\rho^*-e_2}{G_2(1-\rho^*)}-e_4},
\end{split}
\end{equation}
where $e_1$, $e_2$, $e_3$ are defined as in \eqref{EQU-37} and $e_4\triangleq{\frac{\beta G_2 P_{max,2}+ \alpha F_2}{G_2}}$.
\eqref{EQU-75} is further equivalent to
\begin{equation}\label{EQU-76} \vspace{-3pt}
\begin{split}
f(\rho^*)={-e_3(1-\rho^*)+\frac{e_1-e_2}{G_2(1-\rho^*)}-\frac{e_1}{G_2}+e_3-e_4}.
\end{split}
\end{equation}
Hence, problem (32) is simplified as
\begin{equation}\label{EQU-77} \vspace{-3pt}
\begin{split}
&\max_{\rho^*} ~~ f(\rho^*)\\
\textit{s.t.} ~~&P^*_2 \geq {\frac{\frac{{\tau}_1{\bf \sigma}^2_{1,c}}{1-\rho^*}+D_2}{E_2}},\\
&{P^*_1 J_2+P^*_2 K_2}\leq P_r-L_2,\\
&0<P^*_1\leq P_{max,1},\\
&{0<\rho^*<1}.
\end{split}
\end{equation}
Similar to Lemma 3, due to the fact that $e_3>0$, if $\frac{e_1-e_2}{G_2}<0$, i.e., $e_2>e_1$, the objective function $f(\rho^*)$ must have a maximum value, which can be inferred from ${-e_3(1-{\rho^*})}={\frac{e_1-e_2}{G_2(1-\rho^*)}}$. On the other hand, to guarantee the optimal solution ${\rho^*}$ satisfying $0<{\rho^*}<1$, we must have $e_2<e_1+G_2e_3$, which implies that the optimal solution ${\rho^*}$ of problem \eqref{EQU-77} can be derived as
\begin{equation}\label{EQU-78} \vspace{-3pt}
\begin{split}
{\rho^*}=1-\sqrt{\frac{e_2-e_1}{G_2e_3}}.
\end{split}
\end{equation}
Then we obtain the optimal solution in \eqref{EQU-37}.

When the two transmit power constraints \eqref{32d} and \eqref{32e} hold with equality, we obtain the following two equations
\begin{equation}\label{EQU-79} \vspace{-3pt}
P^*_1 = P_{max,1},\quad P^*_2 = {\frac{P_r-L_2-J_2P_{max,1}}{K_2}},
\end{equation}
Based on \eqref{EQU-79}, the two SINR constraints \eqref{32b} and \eqref{32c} can be equivalently written as
\begin{equation}\label{EQU-80} \vspace{-3pt}
{\rho}^* \leq 1-\frac{{\tau}_1 {\bf \sigma}^2_{1,c}}{E_2P^*_2-D_2},
\end{equation}
and
\begin{equation}\label{EQU-81} \vspace{-3pt}
{\rho}^* \leq 1-\frac{{\tau}_2 {\bf \sigma}^2_{2,c}}{G_2P^*_1-F_2}.
\end{equation}
Note that, to guarantee the optimal solution ${\rho^*}$ satisfying $0<{\rho^*}<1$, we must have $0<\frac{{\tau}_1 {\bf \sigma}^2_{1,c}}{E_2P^*_2-D_2}<1$ and $0<\frac{{\tau}_2 {\bf \sigma}^2_{2,c}}{G_2P^*_1-F_2}<1$, which implies that the optimal solution ${\rho^*}$ of problem (32) can be derived as
\begin{equation}\label{EQU-82} \vspace{-3pt}
{\rho^*}=min\{1-\frac{{\tau}_1 {\bf \sigma}^2_{1,c}}{E_2P^*_2-D_2},1-\frac{{\tau}_2 {\bf \sigma}^2_{2,c}}{G_2P^*_1-F_2}\}.
\end{equation}
Then we obtain the optimal solution in \eqref{EQU-38}.

When the constraints \eqref{32d} and \eqref{32f} hold with equality, we obtain the following two equations
\begin{equation}\label{EQU-83} \vspace{-3pt}
P^*_2 = P_{max,2},\quad P^*_1 = {\frac{P_r-L_2-K_2P_{max,2}}{J_2}},
\end{equation}
Then, substituting \eqref{EQU-83} into the two SINR constraints in \eqref{32b} and \eqref{32c}, respectively, which can be equivalently written as
\begin{equation}\label{EQU-84} \vspace{-3pt}
{\rho}^* \leq 1-\frac{{\tau}_1 {\bf \sigma}^2_{1,c}}{E_2P^*_2-D_2},
\end{equation}
and
\begin{equation}\label{EQU-85} \vspace{-3pt}
{\rho}^* \leq 1-\frac{{\tau}_2 {\bf \sigma}^2_{2,c}}{G_2P^*_1-F_2}.
\end{equation}
To guarantee the optimal solution ${\rho^*}$ satisfying $0<{\rho^*}<1$, we must have $0<\frac{{\tau}_1 {\bf \sigma}^2_{1,c}}{E_2P^*_2-D_2}<1$ and $0<\frac{{\tau}_2 {\bf \sigma}^2_{2,c}}{G_2P^*_1-F_2}<1$, which implies that the optimal solution ${\rho^*}$ of problem (32) can be derived as
\begin{equation}\label{EQU-86} \vspace{-3pt}
{\rho^*}=min\{1-\frac{{\tau}_1 {\bf \sigma}^2_{1,c}}{E_2P^*_2-D_2},1-\frac{{\tau}_2 {\bf \sigma}^2_{2,c}}{G_2P^*_1-F_2}\}.
\end{equation}
Then we obtain the optimal solution in \eqref{EQU-39}.

When the constraints \eqref{32e} and \eqref{32f} hold with equality, we obtain the following two equations
\begin{equation}\label{EQU-87} \vspace{-3pt}
P^*_1 = P_{max,1}, \quad P^*_2 = P_{max,2},
\end{equation}
Hence, the two SINR constraints \eqref{32b} and \eqref{32c} can be equivalently written as
\begin{equation}\label{EQU-88} \vspace{-3pt}
{\rho}^* \leq 1-\frac{{\tau}_1 {\bf \sigma}^2_{1,c}}{E_2P^*_2-D_2},
\end{equation}
and
\begin{equation}\label{EQU-89} \vspace{-3pt}
{\rho}^* \leq 1-\frac{{\tau}_2 {\bf \sigma}^2_{2,c}}{G_2P^*_1-F_2}.
\end{equation}
Similar to above discussion, the optimal solution ${\rho^*}$ of problem (32) can be derived as
\begin{equation}\label{EQU-90} \vspace{-3pt}
{\rho^*}=min\{1-\frac{{\tau}_1 {\bf \sigma}^2_{1,c}}{E_2P^*_2-D_2},1-\frac{{\tau}_2 {\bf \sigma}^2_{2,c}}{G_2P^*_1-F_2}\}.
\end{equation}
Then we obtain the optimal solution in \eqref{EQU-40}.
The proof of Lemma 2 is thus completed.\\

\section{Proof of Lemma 3}
First, the objective function $f(\rho^*)$ in problem \eqref{EQU-58} is equivalently written as
\begin{equation}\label{EQU-91} \vspace{-3pt}
\begin{split}
{f(\rho^*)}=&{\frac{a_1({(\rho^*)}^2-1)+a_2(\rho^*-1)+a_1+a_2-a_3}{E_2G_2(1-\rho^*)}-a_4},\\
=&{\frac{a_1(1-{\rho^*})}{E_2G_2}}+{\frac{a_1+a_2-a_3}{E_2G_2(1-\rho^*)}}+{\frac{2a_1-a_2}{E_2G_2}}-a_4.
\end{split}
\end{equation}
According to the property of the function $f(x)={ax+\frac{b}{x}}$, the objective function $f(\rho^*)$ have a maximum value when ${\frac{a_1}{E_2G_2}}<0$ and ${\frac{a_1+a_2-a_3}{E_2G_2}}<0$. Due to the fact that $a_1<0$ and $E_2G_2>0$, we have ${\frac{a_1}{E_2G_2}}<0$. Hence, if $a_1+a_2-a_3<0$, i.e., $a_2-a_3<-a_1$, the objective function $f(\rho^*)$ must exist the maximum value, which can be inferred from the fact ${\frac{a_1(1-{\rho^*})}{E_2G_2}}={\frac{a_1+a_2-a_3}{E_2G_2(1-\rho^*)}}$. Note that, to guarantee the optimal solution ${\rho^*}$ satisfying $0<{\rho^*}<1$, we also let $a_2-a_3>0$. As a result, the optimal solution ${\rho^*}$ of problem \eqref{EQU-58} can be derived as
\begin{equation}\label{EQU-92} \vspace{-3pt}
\begin{split}
{\rho^*}=1-\sqrt{\frac{a_1+a_2-a_3}{a_1}},
\end{split}
\end{equation}
Next, we show that $a_1+a_2-a_3>0$ cannot happen at the optimal solution ${\rho^*}$. We prove this result by contradiction. In this case, if we want the objective function $f(\rho^*)$ to increase in \eqref{EQU-91}, the optimal solution ${\rho^*}$ will be ${\rho^*}\rightarrow1$, which leads to that the transmit power solution $P^*_i\rightarrow\infty$. It is easy to verify that the above case cannot happen due to the transmit power constraints in problem \eqref{EQU-58}. In conclusion, the optimal solution ${\rho^*}$ of \eqref{EQU-58} can be obtained while $0<a_2-a_3<-a_1$. The proof of Lemma 3 is thus completed.\\



\end{document}